\documentclass[a4paper,11pt]{article}
\usepackage{jheppub} 
\usepackage{lineno}
\usepackage{amsmath}
\usepackage{multirow}
\usepackage{slashed}
\usepackage{orcidlink}

\usepackage{tikz}
\usepackage{tikz-feynman}
\usepackage{caption}
\usepackage{subcaption}
\usepackage{graphicx}      
\usepackage{caption}       
\usepackage{float}         
\usepackage{xcolor}
\usepackage{hyperref}

\preprint{IPPP/26/06, IOP/BBSR/2026-02}

\newcommand{\opQ}[1]{\mathcal{O}_{#1}}
\newcommand{\opq}[2]{$\mathcal{O}_{#1}^{(#2)}$}
\newcommand{\hc}{\text{h.c.}}

\newcommand{\blv}{}

\newcommand{\blvmc}[1]{\multicolumn{3}{c}{#1}}

\newcommand{\nsmeft}{\nu\text{SMEFT}}

\title{Differentiating Dimension-6 and Dimension-8 Effects in $\nsmeft$ at the HL-LHC}

\abstract{
We study dimension-eight effects in the Standard Model Effective Field Theory extended by right-handed neutrinos ($\nsmeft$). Using the Hilbert series formalism, we derive the complete basis of dimension-eight operators and confirm agreement with existing classifications, providing a systematic framework beyond the conventional dimension-six truncation.
We analyse the collider phenomenology of the representative operator $\opQ{N^{2}q^{2}B}^{(1,2)}$ at the High-Luminosity LHC. The resulting signatures involve pair production of right-handed neutrinos in association with jets, followed by decays into electron-jet final states with potentially displaced vertices. Since similar final states are generated by leading dimension-six operators, we explicitly address whether dimension-eight contributions can be experimentally distinguished from dimension-six effects.
Using a Boosted Decision Tree analysis based on kinematic observables, we show that the dimension-eight signal can be reliably separated from each relevant dimension-six hypothesis. Our results demonstrate that dimension-eight operators in the $\nsmeft$ can give rise to experimentally resolvable signatures and should be included in collider EFT interpretations.
}

\author[a,b]{Manimala Mitra\orcidlink{0000-0002-8032-5125},} 
\author[c]{Shakeel Ur Rahaman\orcidlink{0000-0001-6153-8187},}
\author[a,b]{Subham Saha\orcidlink{0009-0009-1183-3271},}
\author[c]{Michael Spannowsky\orcidlink{0000-0002-8362-0576}}
\affiliation[a]{Institute of Physics, Sachivalaya Marg, Bhubaneswar, Odisha 751005, India}
\affiliation[b]{Homi Bhabha National Institute, BARC Training School Complex, Anushakti Nagar, Mumbai, 400094, India}
\affiliation[c]{Institute for Particle Physics Phenomenology, Department of Physics, Durham University,\\South Road, Durham DH1 3LE, United Kingdom}
\emailAdd{manimala@iopb.res.in}
\emailAdd{shakeel.u.rahaman@durham.ac.uk}
\emailAdd{subham.saha@iopb.res.in}
\emailAdd{michael.spannowsky@durham.ac.uk}

\begin{document}
\maketitle

\section{Introduction}\label{sec:intro}

The Standard Model (SM) of particle physics has achieved remarkable success in describing a wide range of experimental data. Nevertheless, it fails to account for the observed neutrino masses and mixings, which have been firmly established by neutrino oscillation experiments \cite{Esteban:2020cvm,deSalas:2020pgw}. This provides clear evidence for physics beyond the SM (BSM). The Type I seesaw mechanism~\cite{Minkowski:1977sc,Gell-Mann:1979vob,Mohapatra:1979ia,Yanagida:1979as,Yanagida:1980xy}, that extends the SM with heavy right-handed neutrinos (RHNs) with a Majorana-like mass term, stands as one of the most compelling frameworks for explaining such a scenario.

Motivated by collider phenomenology, a variety of low-scale seesaw mechanisms have been proposed, including inverse \cite{Mohapatra:1986aw,Mohapatra:1986bd} and linear \cite{Akhmedov:1995ip,Malinsky:2005bi} seesaws, that predict the RHNs at experimentally accessible energies. In this kind of scenario, where the RHNs are light, they cannot be integrated out at the electroweak (EW) scale. The effective field theory (EFT) required to describe physics beyond the EW scale must also be built, including the RHN sector coupled to SM particles. Such an EFT is known as $\nsmeft$. Constructing the $\nsmeft$ operator basis is essential for systematic phenomenological studies but poses a significant challenge. Recent advancements in the Hilbert Series (HS) method have addressed this issue \cite{DELAGUILA2009399,PhysRevD.80.013010,PhysRevD.94.055022,PhysRevD.96.015012,Li:2021tsq,Alonso:2024usj,Banerjee:2020bym,Banerjee:2020jun,Graf:2022rco,Hanany:2010vu,Henning:2015alf,Henning:2017fpj,Lehman:2015via,Lehman:2015coa}. By exploiting group-theoretic techniques, it provides a systematic classification of all independent operators at a given mass dimension, taking into account gauge symmetries and redundancies arising from equations of motion and integration by parts. We adopt the methodology for enumerating operators at mass dimension-8 (D8) for $\nsmeft$. 

Collider phenomenology associated with lower-dimensional operators in this construction, particularly those of dimension-5 (D5) and dimension-6 (D6), has been extensively studied at the LHC and future colliders \cite{Englert:2015hrx,Caputo:2017pit,Barducci:2020icf,Delgado:2022fea,Duarte:2023tdw,Jones-Perez:2019plk, Mitra:2024ebr}. Whereas the collider implications of D8 operators in the $\nsmeft$ have remained comparatively unexplored \cite{Banerjee:2022thk}. At high energies, such contributions could become phenomenologically relevant due to their enhanced energy dependence, potentially leading to observable deviations from expectations based solely on D6 operators. Furthermore, because effective operators are implemented at the amplitude level, a double insertion of a D6 operator carries the same suppression as a single D8 operator insertion. Developing a strategy to distinguish these two effects is therefore essential for probing the underlying UV-completion and may ultimately help differentiate among competing UV models~\cite{Naskar:2022rpg, Boughezal:2021tih, Beltran:2023ymm}.

In this work, we construct the D8 operator basis and cross-check the counting against existing literature. We then investigate the collider phenomenology of a specific D8 operator, $\opQ{N^2 q^2 B}^{(1)}$, at the High-Luminosity LHC (HL-LHC). Our primary objective is to assess the extent to which the effects of this operator can be distinguished from those induced by leading D6 operators, specifically $\opQ{qN}$ and $\opQ{duNe}$. We focus on single RHN production via Drell-Yan processes, followed by the leptonic decay $N \to e \ell \nu$, which yields a trilepton final state with missing transverse energy. This D8 operator introduces a new topology for this process; depending on the RHN lifetime, such scenarios may also produce characteristic displaced vertex signatures \cite{Drewes:2019fou,Cottin:2021lrq,Liu:2019ayx,Abada:2018sfh}.

The remainder of this paper is organised as follows. In Section~\ref{sec:theory}, we outline the theoretical framework, briefly introducing the Hilbert series method used to derive the D8 $\nsmeft$ operator basis and presenting the operators relevant to our study. Section~\ref{sec:prod-dk-an} details the RHN production cross-sections and decay modes, while also describing the collider analysis and BDT-based discrimination strategy at the HL-LHC. In Section~\ref{sec:results}, we present our numerical results and discuss current constraints on the Wilson coefficients. Finally, we summarise our findings and provide concluding remarks in Section~\ref{sec:conclusion}.

\section{Theoretical Framework} \label{sec:theory}

Existing literature has already identified the $\nsmeft$ dimension 8 operators employing some other methods \cite{Li:2021tsq}\footnote{To keep things concise, we only write the operators that involve the RHN.}. General discussion of operator construction within the HS framework and a comparison with the existing literature will be deferred to Appendix~\ref{app:op}. This section will therefore concentrate on the operators relevant to our specific phenomenological study and the introduction of the notations used.

\subsection{Notation and Operators}
The $\nsmeft$ Lagrangian is invariant under $SU(3)_{C} \times SU(2)_{w} \times U(1)_{Y}$ internal symmetry group, and the spacetime symmetry is governed by the Lorentz group. The fields that constitute the operators and their transformation properties are shown in Table~\ref{tab:vSMEFT-field-content}.

\begin{table}[h]
\centering
    \renewcommand{\arraystretch}{1.5}
    {\begin{tabular}{c|ccccc|c}
    & & $SU(3)_{C}$ & $SU(2)_{w}$ & $U(1)_{Y}$ & $SU(2)_{L}\times SU(2)_{R}$ & \# flavors\\
    \hline
    $\mathbf{H}$& $H$ & $\boldsymbol{1}$ & $\boldsymbol{2}$ & $\frac{1}{2}$ & $\left(0,0\right)$ & $1$\\
    \hline
    \multirow{6}{*}{$\mathbf{\Psi}$}& $q_L$ & $\boldsymbol{3}$ & $\boldsymbol{2}$ & $\frac{1}{6}$ & $\left(\frac{1}{2},0\right)$ & $n_q$\\
    & $u_R$ & $\boldsymbol{3}$ & $\boldsymbol{1}$ & $\frac{2}{3}$ & $\left(0,\frac{1}{2}\right)$ & $n_u$\\
    & $d_{R}$ & $\boldsymbol{3}$ & $\boldsymbol{1}$ & $-\frac{1}{3}$ & $\left(0,\frac{1}{2}\right)$ & $n_d$\\
    & $l_L$ & $\boldsymbol{1}$ & $\boldsymbol{2}$ & $-\frac{1}{2}$ & $\left(\frac{1}{2},0\right)$ & $n_l$\\
    & $e_{R}$ & $\boldsymbol{1}$ & $\boldsymbol{1}$ & $-1$ & $\left(0,\frac{1}{2}\right)$ & $n_e$\\
    & $N_{R}$ & $\boldsymbol{1}$ & $\boldsymbol{1}$ & $0$ & $\left(0,\frac{1}{2}\right)$ & $n_N$\\
    \hline
    \multirow{3}{*}{$\mathbf{X}$}& $G_L$ & $\boldsymbol{8}$ & $\boldsymbol{1}$ & 0 & $\left(1,0\right)$ & $1$\\
    & $W_L$ & $\boldsymbol{1}$ & $\boldsymbol{3}$ & 0 & $\left(1,0\right)$ & $1$\\
    & $B_L$ & $\boldsymbol{1}$ & $\boldsymbol{1}$ & 0 & $\left(1,0\right)$ & $1$\\
    \hline
    $\mathbf{D}$ & $D$ & \multicolumn{3}{c}{\textbf{Covariant Derivative}} &  $\left( \frac12, \frac12 \right)$ & \\
    \hline
    \end{tabular}}
\caption{\label{tab:vSMEFT-field-content}$\nu$SMEFT field content. }
\end{table}
The leftmost column denotes the generic variable for the different field spins. $\mathbf{H}$ denotes the scalar field, $\mathbf{\Psi}$ denotes the fermionic field, $\mathbf{X}$ and $\mathbf{D}$ denotes the Field strength tensor and covariant derivative respectively. To demarcate the operators into various operator classes, we shall use those variables. The Lorentz group $SO(3,1)$ is decomposed to $SU(2)_{L}\times SU(2)_{R}$, where the subscript $L$ and $R$ denotes left handed and right handed component. 

Before we delve into the long list of operators, we want to make a few points.
\begin{itemize}
    \item We are using the term ``operator" in a broad sense, without distinguishing between different types of operators and the Lagrangian terms. This approach will be clarified later and is discussed in detail in ref.~\cite{Fonseca:2019yya}. Accordingly, we will only tabulate the ``types of terms", and we will not explicitly write out their hermitian conjugates.
	\item Within the parenthesis the indices for various groups are usually contracted and implicit. If they are not contracted, we have explicitly mentioned them. We use $\{i,j,k,l\} \in \{1,2\}$ indices for $SU(2)$ fundamental representation,
    $\{\alpha,\beta,\gamma\} \in \{1,2,3\}$ indices for $SU(3)$ fundamental. $\{A,B,C\} \in \{1,2,...,8\}$ are used for $SU(3)$, $\{I,J,K\} \in \{1,2,3\}$ for $SU(2)$ adjoint represenation. $\{ \mu, \nu,\rho,\lambda\}$ are the spacetime indices.
    \item We have also omitted the flavour indices $(p,q,r,s)$ in writing the operator structure, but have included their contributions when counting.
    \item We are also going to suppress the chiral subscript $L/R$, it shall be clear from the notation of the fields defined in Table~\ref{tab:vSMEFT-field-content}. 
    
\end{itemize}
Our analysis will be restricted to the operators that give rise to new interaction vertices, which cannot be generated at D6 and only first appear at D8 or higher. Furthermore, we exclude operators that violate Baryon number ($\slashed{B}$) or Lepton number ($\slashed{L}$), his restriction naturally limits our focus to even-dimensional operators~\cite{Kobach:2016ami}. Keeping this in mind, we concentrate on the $\Psi^4 X$ class as a representative and phenomenologically well-motivated example. A complete basis of operators belonging to this class is listed in Table~\ref{tab:dim8-ops-psi4x-1}. 
\begin{table}[h]
    \centering
    \renewcommand{\arraystretch}{1.5}
    {\begin{tabular}{c|c|c|c}
        \multicolumn{4}{c}{$\mathbf{\Psi^4 X }\;\;(+\hc)$}\\
        \hline
        \hline
        $\opQ{d^2N^2B}$&
        $ (\bar{N} \gamma^{\mu} N)(\bar{d} \gamma^{\nu} d)B_{\mu \nu} $ &
        $\opQ{d^2N^2G}$&
        $(\bar{N} \gamma^{\mu} N)(\bar{d} \gamma^{\nu} T^{A} d)G_{\mu \nu}^{A} $ \\
        $\opQ{e^2N^2B}$&
        $ (\bar{N} \gamma^{\mu} N)(\bar{e} \gamma^{\nu} e)B_{\mu \nu} $ &
        $\opQ{N^4B}$&
        $ (\bar{N} \gamma^{\mu} N)(\bar{N} \gamma^{\nu} N)B_{\mu \nu} $ \\
        $\opQ{N^2u^2B}$&
        $ (\bar{N} \gamma^{\mu} N)(\bar{u} \gamma^{\nu} u)B_{\mu \nu} $ &
        $\opQ{N^2u^2G}$&
        $ (\bar{N} \gamma^{\mu} N)(\bar{u} \gamma^{\nu} T^{A} u)G_{\mu \nu}^{A} $ \\
        $\opQ{l^2N^2B}$&
        $  (\bar{l} \gamma^{\mu} l)(\bar{N} \gamma^{\nu} N) B_{\mu \nu} $ &
        $\opQ{l^2N^2W}$&
        $ (\bar{l} \gamma^{\mu} \tau^{I} l)(\bar{N} \gamma^{\nu} N) W_{\mu \nu}^{I}  $ \\
        $\opQ{N^2q^2B}$&
        $  (\bar{q} \gamma^{\mu} q)(\bar{N} \gamma^{\nu} N) B_{\mu \nu} $ &
        $\opQ{N^2q^2W}$&
        $ (\bar{q} \gamma^{\mu} \tau^{I} q)(\bar{N} \gamma^{\nu} N) W_{\mu \nu}^{I} $ \\ 
        $\opQ{N^2q^2G}$&
        $ (\bar{q} \gamma^{\mu} T^{A} q)(\bar{N} \gamma^{\nu} N) G_{\mu \nu}^{A} $ &
        $\opQ{el^2NB} $& 
        $(\bar{l^{j}} \sigma^{\mu \nu} e) \epsilon_{jk} (\bar{l^{k}} N) B_{\mu \nu} $ \\

        $ \opQ{el^2NW}$ &
        $ (\bar{l^{j}} \sigma^{\mu \nu} N) {(\tau^{I} \epsilon)}_{jk} (\bar{l^{k}} e) W_{\mu \nu}^{I} $ &
        $ \opQ{dlNqW}$ &
        $ (\bar{l^{j}} \sigma^{ \mu \nu} N) {(\tau^{I} \epsilon)}_{jk} (\bar{q^{k}}d) W_{\mu \nu}^{I} $ \\
        
        $ \opQ{dlNqB} $ &
        $ (\bar{l^{j}} \sigma^{\mu \nu} N) \epsilon_{jk} (\bar{q^{k}} d) B_{\mu \nu} $ &
        $ \opQ{dlNqG} $ &
        $  (\bar{l^{j}} \sigma^{\mu \nu} N) \epsilon_{jk} (\bar{q^{k}} T^{A} d) G_{\mu \nu}^{A} 
        $ \\
        \opq{deNuB}{1}  &
        $  (\bar{u} \gamma^{\mu} d) (\bar{e} \gamma^{\nu} N) B_{\mu \nu} $ &
        \opq{deNuB}{2} &
        $  (\bar{u} \gamma^{\mu} d) (\bar{e} \gamma^{\nu} N) \widetilde{B}_{\mu \nu} $ \\
        
        \opq{deNuG}{1} &
        $ (\bar{u} T^{A} \gamma^{\mu} d) (\bar{e} \gamma^{\nu}  N) G_{\mu \nu}^{A} $ &
        \opq{deNuG}{2} &
        $ (\bar{u} T^{A} \gamma^{\mu} d) (\bar{e} \gamma^{\nu} N) \widetilde{G}_{\mu \nu}^{A} $ \\
        
        \opq{lNquB}{1} &      
        $ (\bar{l^{j}} \sigma^{\mu \nu} N) (\bar{u} q_{j}) B_{\mu \nu} $ &  \opq{lNquB}{2} &      
        $ (\bar{l^{j}} N) (\bar{u} \sigma^{\mu \nu} q_{j}) \widetilde{B}_{\mu \nu} $ \\

        \opq{lNquG}{1} &
        $ (\bar{l^{j}} \sigma^{\mu \nu} N) (\bar{u} T^{A} q_{j}) G_{\mu \nu}^{A} $ &
        \opq{lNquG}{2} &
        $ (\bar{l^{j}} N) (\bar{u} \sigma^{\mu \nu} T^{A} q_{j}) \widetilde{G}_{\mu \nu}^{A} $ \\
        
        \opq{lNquW}{1} &
        $ (\bar{l^{j}} \sigma^{\mu \nu} N) (\tau^{I})_{jk} (\bar{u} q^{k}) W_{\mu \nu}^{I} $ &
        \opq{lNquW}{2} &
        $ (\bar{l^{j}} N) (\tau^{I})_{jk} (\bar{u} \sigma^{\mu \nu} q^{k}) \widetilde{W}_{\mu \nu}^{I} $ \\
        \hline
        \end{tabular}}
    \caption{Operators in $\mathbf{\Psi^4 X }$ class that conserves $B, L$-numbers. All the operators here have their hermitian conjugates. The rest of the operators that violate baryon and lepton number $(\slashed{B},\slashed{L})$ are presented in Table~\ref{tab:dim8-ops-psi4x-2}. }
	\label{tab:dim8-ops-psi4x-1}
\end{table}
From this set, we single out the operators that mediate the partonic process $pp \to NNjj$ via an intermediate $ Z$-boson. In particular, the operators \opq{N^2q^2B}{1} and \opq{N^2q^2B}{2} are identified as the relevant ones for our signal topology.\footnote{We present only the operator types; fixing the number of fermion generations to one yields two independent operator contractions for this specific type.}

\subsection{Relevant Higher Order Interactions}

The RHN mixes with the SM electron neutrino ($\nu_e$) via a small active-sterile mixing, denoted by $\chi$. At the renormalisable level, the interaction between the RHN and SM fields is expressed as 
\begin{equation}
\mathcal{L}_{lNW} = \frac{g}{\sqrt{2}} W^-_{\mu} \bar{e} \gamma^{\mu} P_L \chi N + \hc\;.
\label{eq:renormNRint}
\end{equation}
To simplify the analysis, we consider only one lepton flavour within the charged-current interactions; this can be easily extended to other lepton flavours. Furthermore, the RHN participates in neutral-current interactions with the $Z$ boson and Yukawa-type interactions with the Higgs boson, both of which are mediated by its mixing with the active neutrinos. The higher-dimensional terms (D6 and D8), relevant to our study in the Lagrangian, can be schematically written as,
\begin{align}
    \mathcal{L} \subset \frac{C_{qN}}{\Lambda^2} \opQ{qN}+ \frac{C_{duNe}}{\Lambda^2} \opQ{duNe} + \frac{C_{N^2 q^2 B}^{(1)}}{\Lambda^4} \opQ{N^2 q^2 B}^{(1)} + \frac{C_{N^2 q^2 B}^{(2)}}{\Lambda^4} \opQ{N^2 q^2 B}^{(2)},
\end{align}
where the D6 operators are defined as\footnote{See Refs.~\cite{Banerjee:2015gca,Mitra:2022nri}, for comprehensive studies of these operators.},
\begin{align} \label{eq:duNe}
\opQ{qN}=\left(\bar{q}\gamma^{\mu}q\right)\left(\bar{N}\gamma_{\mu}N\right)\,; \quad
\opQ{duNe}=\left(\bar{d}\gamma^{\mu}u\right)\left(\bar{N}\gamma_{\mu}e\right) + \hc \,,
\end{align}
the D8 operators are defined as,
\begin{align}\label{eq:ONNBB}
 \mathcal{O}^{(1)}_{N^2 q^2 B} =(\bar{q} \gamma^{\mu} q)(\bar{N} \gamma^{\nu} N) {B}_{\mu \nu}\;; \quad \mathcal{O}^{(2)}_{N^2 q^2 B} = (\bar{q} \gamma^{\mu} q)(\bar{N} \gamma^{\nu} N) \tilde{B}_{\mu \nu}\,,
\end{align}
and the coefficients $C_i$ are the Wilson coefficients (WCs) scaled appropriately by the new physics scale ($\Lambda$). The model parameters, including the WCs, are chosen so that the RHN decays are displaced, thereby rendering the SM background effectively negligible. Since the described D6 operators can induce similar final states, we focus on distinguishing their contributions from those of D8 operators. To this end, the next section provides a quantitative comparison using a multivariate analysis.

\section{Production, Decay, and Analysis} \label{sec:prod-dk-an}
\subsection{Decay} \label{sec:decay}

The allowed decay modes of the RHN are constrained by the specific structure of the higher-dimensional operators under consideration. Notably, since $\opQ{qN}$ and $\opQ{N^2 q^2 B}^{(1,2)}$ involve at least two RHN fields, they do not facilitate the decay of a single $N$ into Standard Model states. In such cases, the RHN decay proceeds through the active-sterile mixing $\chi$. Conversely, the introduction of the $\opQ{duNe}$ operator provides a direct, contact interaction channel for the $N \to e j j$ decay.

The decay patterns of the RHN depend crucially on its mass. For $M_N < M_W$, the allowed channels are $N \to e \ell^{\prime} \nu$ (mediated by $W$), $N \to e j j$ (mediated by $W$ and $C_{duNe}$), $N \to \nu_e j j$ and $N \to \nu_e \nu \nu$ (both mediated by $Z$), as well as $N \to \nu_e \ell^{\prime +} \ell^{\prime -}$, $N \to \nu_e c \bar{c}$, and $N \to \nu_e b \bar{b}$ (mediated by $Z$ and $H$). For $M_N > M_W$, the two-body modes $N \to e W$ and $N \to e j j$ become kinematically allowed. When $M_N > M_Z$, additional channels such as $N \to \nu_e Z$ open up, while for $M_N > M_H$, the decay $N \to \nu_e H$ is also possible alongside the previously mentioned modes. Here, $j$ denotes light quarks and $\ell^{\prime}$ represents $e$, $\mu$, or $\tau$.  Among these channels, the $e j j$ decay mode receives a significant contribution from $C_{duNe}$. For sufficiently large values of this coupling, the $e j j$ branching ratio can even dominate over the standard two-body decays. For intermediate mass ranges such as $M_W < M_N < M_Z$, off-shell processes like $N \to \nu_e Z^* \to \nu_e f \bar{f}$, where $f$ denotes SM fermions, are possible but remain suppressed compared to $N \to e W$ and $N \to e j j$. A similar suppression applies to Higgs-mediated three-body decays when $M_Z < M_N < M_H$. The expressions for the partial decay widths are taken from the Appendix of Ref.~\cite{Mitra:2024ebr}.  
 
\begin{figure}[h]
\centering
\begin{minipage}[b]{0.45\textwidth}
    \centering
    \includegraphics[scale=0.45]{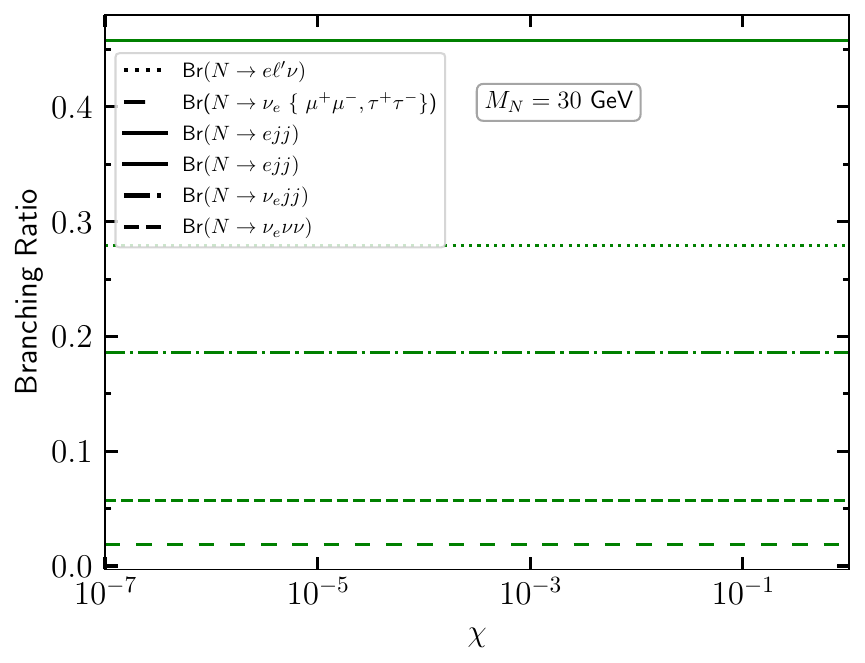}
\end{minipage}
\hfill
\begin{minipage}[b]{0.45\textwidth}
    \centering
    \includegraphics[scale=0.45]{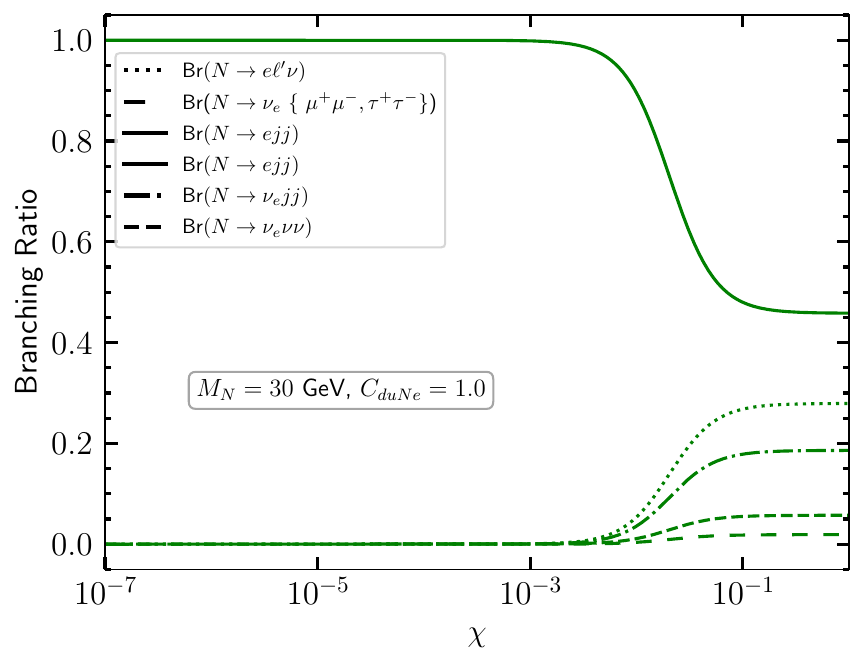}
\end{minipage}

\vspace{0.5cm} 

\begin{minipage}[b]{0.45\textwidth}
    \centering
    \includegraphics[scale=0.45]{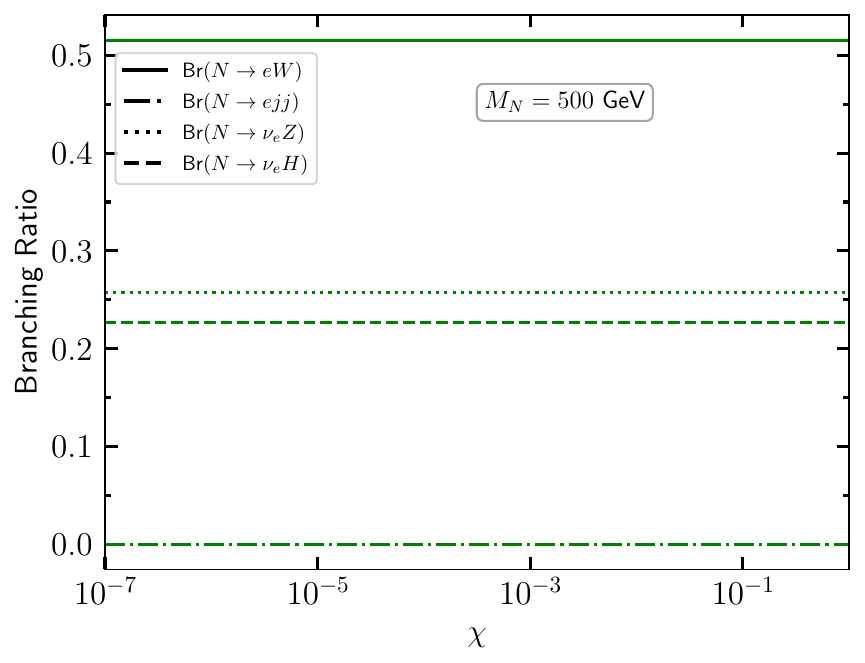}
\end{minipage}
\hfill
\begin{minipage}[b]{0.45\textwidth}
    \centering
    \includegraphics[scale=0.45]{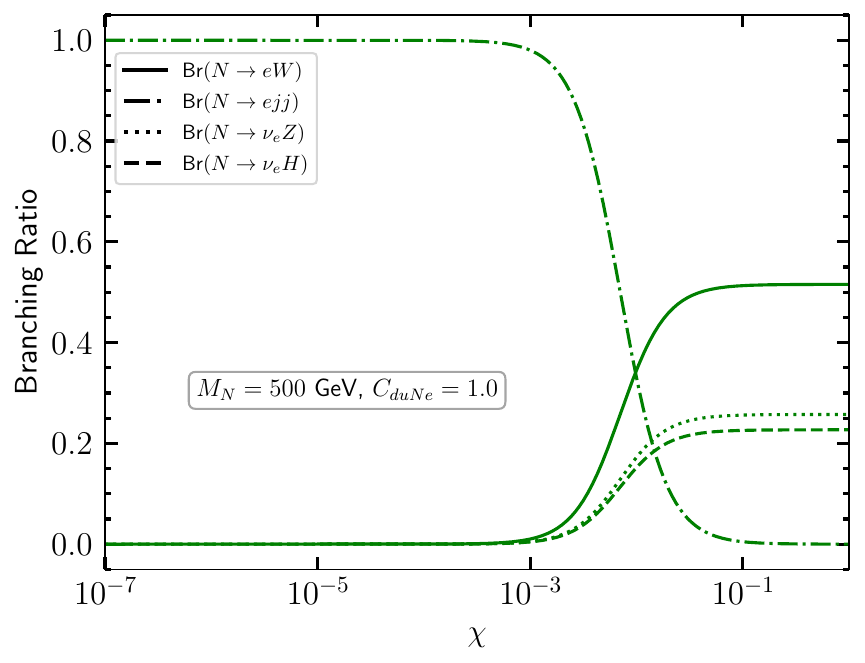}
\end{minipage}

\caption{Branching ratios of the RHN with variation of $\chi$ (active-sterile neutrino mixing). \textbf{Top row:} $M_N=30$ GeV for $C_{duNe}=0$  (left) and $C_{duNe}=1$  (right). \textbf{Bottom row:} $M_N=500$ GeV for $C_{duNe}=0$  (left) and $C_{duNe}=1$  (right).} 
\label{fig:BR_combined}
\end{figure}

In Fig.~\ref{fig:BR_combined}, we show the variation of the branching ratios as a function of the mixing parameter $\chi$. The upper panels correspond to the case with $M_N = 30~\text{GeV}$, where the left panel illustrates the scenario including only $\opQ{qN}$ or $\opQ{N^2 q^2 B}^{(1,2)}$, while the right panel displays the case with only the $\opQ{duNe}$ operator. The lower panels present the corresponding results for $M_N = 500~\text{GeV}$. A few important observations can be made from Fig.~\ref{fig:BR_combined}. When the decay of $N$ proceeds only through the mixing parameter $\chi$, all decay channels exhibit the same $\chi$-dependence in their partial widths. Consequently, this dependence cancels in the branching ratios, resulting in a flat behaviour, as seen in the left panels of the figure.  

For the case $M_N = 30~\text{GeV}$, the decay mode $N \to ejj$ dominates, while for $M_N = 500~\text{GeV}$ the two-body mode $N \to eW$ becomes dominant. The latter subsequently leads to $ejj$ final states through the on-shell decay of the $W$ boson. In the absence of the $C_{duNe}$ operator, the three-body $ejj$ mode vanishes for $M_N = 500~\text{GeV}$, which is consistent with expectations.  

When the $\opQ{duNe}$ operator is included, the behaviour changes significantly. For $M_N = 30~\text{GeV}$, the branching ratio of $N \to ejj$ remains unity up to $\chi \simeq 10^{-3}$ due to the dominance of the $C_{duNe}$ contribution. Beyond this point, there is a competition between the $ejj$ channel mediated by the off-shell $W$ boson and the $C_{duNe}$ contribution, as well as other decay modes mediated by $\chi$. As $\chi$ increases, these channels gradually saturate, leading to a stable branching pattern. A similar trend is observed for $M_N = 500~\text{GeV}$; however, in this case, the branching ratio of $N \to ejj$ saturates to zero at large $\chi$, since this channel is no longer supported by the $C_{duNe}$ operator alone.

In the subsequent analysis, we focus on the decay channel \(N \to e j j\) and fix the mixing parameter to \(\chi = 10^{-5}\) for \(M_N = 30~\text{GeV}\) and to \(\chi = 10^{-8}\) for \(M_N = 500~\text{GeV}\). For these benchmark mass choices, the corresponding branching ratios are found to be \(\text{BR}(N \to e j j) = 0.46\) and \(0.35\) for \(M_N = 30~\text{GeV}\) and \(500~\text{GeV}\), respectively, when \(C_{duNe} = 0\). Increasing the Wilson coefficient to \(C_{duNe} = 1\) enhances the branching fraction to \(\text{BR}(N \to e j j) \simeq 0.99\) for both mass benchmarks.

\subsection{Production Cross Section}
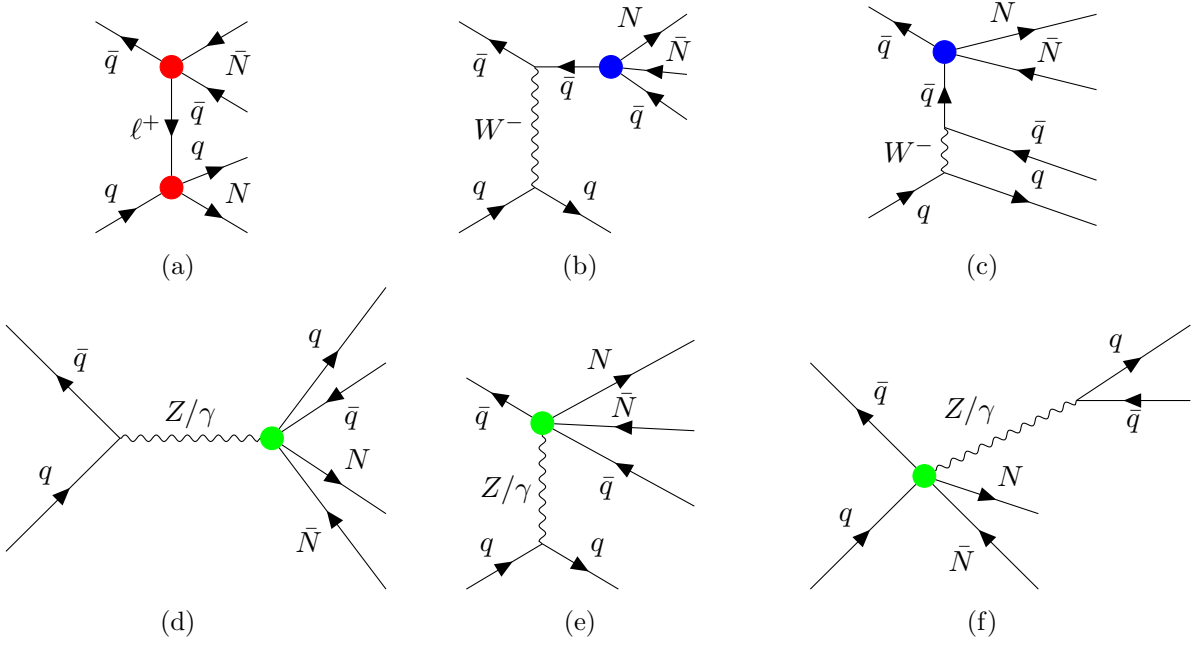
\begin{figure}[h!]
    \centering
    \begin{subfigure}[b]{0.30\textwidth}
        \centering
   \begin{tikzpicture}
  \begin{feynman}
    \vertex[blob, draw=red, fill=red, minimum size=3mm] (a) at (0,0) {};
    \vertex (b) at (-1.0, 0.6);
    \vertex (c) at (1.0, 0.6);
    \vertex (g) at (1.0, -0.6);
    \vertex[blob, draw=red, fill=red, minimum size=3mm] (f) at (0,-1.6) {};
    \vertex (d) at (1.0, -2.2);
    \vertex (h) at (1.0, -1.2);
    \vertex (e) at (-1.0, -2.2);

    \diagram* {
      (a) -- [edge label=$\bar{q}$, fermion] (b),
      (c) -- [edge label=$\bar{N}$, fermion] (a),
      (g) -- [edge label=$\bar{q}$, fermion] (a),
      (a) -- [edge label'=$\ell^{+}$, fermion] (f),
      (f) -- [edge label=$N$, fermion] (d),
      (f) -- [edge label=$q$, fermion] (h),
      (e) -- [edge label=$q$, fermion] (f),
    };
  \end{feynman}
\end{tikzpicture}
    \caption{}
    \label{fig:DuNe}    
    \end{subfigure}
\hfill
\begin{subfigure}[b]{0.30\textwidth}
        \centering
\begin{tikzpicture}
  \begin{feynman}
    \vertex (a) at (0,0);
    \vertex (b) at (-1.0, 0.6);
    \vertex[blob, draw=blue, fill=blue, minimum size=3mm] (c) at (1.0, 0) {};
    \vertex (g) at (2.0, -0.7);
    \vertex (m) at (2.0, -0.1);
    \vertex (n) at (2.0, 0.7);
    \vertex (f) at (0,-1.6);
    \vertex (d) at (1.0, -2.2);
    \vertex (h) at (1.0, -1.2);
    \vertex (e) at (-1.0, -2.2);

    \diagram* {
      (a) -- [edge label=$\bar{q}$, fermion] (b),
      (c) -- [edge label=$\bar{q}$, fermion] (a),
      (c) -- [edge label=$N$, fermion] (n),
      (m) -- [edge label'=$\bar{N}$, fermion] (c),
      (g) -- [edge label=$\bar{q}$, fermion] (c),
      (a) -- [edge label'=$W^{-}$, boson] (f),
      (f) -- [edge label=$q$, fermion] (d),
      (e) -- [edge label=$q$, fermion] (f),
    };
  \end{feynman}
\end{tikzpicture}
    \caption{}
    \label{fig:QN1}    
    \end{subfigure}
\hfill
\begin{subfigure}[b]{0.30\textwidth}
        \centering
\begin{tikzpicture}
  \begin{feynman}
    \vertex[blob, draw=blue, fill=blue, minimum size=3mm] (a) at (0,0) {};
    \vertex (b) at (-1.0, 0.6);
    \vertex (c) at (1.0, 0);
    \vertex (g) at (0, -1.0);
    \vertex (m) at (2.0, -0.5);
    \vertex (n) at (2.0, 0.5);
    \vertex (f) at (0,-1.6);
    \vertex (d) at (2.0, -2.3);
    \vertex (h) at (2.0, -1.7);
    \vertex (e) at (-1.0, -2.2);

    \diagram* {
      (a) -- [edge label=$\bar{q}$, fermion] (b),
      (a) -- [edge label=$N$, fermion] (n),
      (m) -- [edge label'=$\bar{N}$, fermion] (a),
      (g) -- [edge label=$\bar{q}$, fermion] (a),
      (g) -- [edge label'=$W^{-}$, boson] (f),
      (h) -- [edge label'=$\bar{q}$, fermion] (g),
      (f) -- [edge label=$q$, fermion] (d),
      (e) -- [edge label'=$q$, fermion] (f),
    };
  \end{feynman}
\end{tikzpicture}
    \caption{}
    \label{fig:QN2}    
    \end{subfigure}
\vspace{1.0cm}
    \begin{subfigure}[b]{0.30\textwidth}
        \centering
  \begin{tikzpicture}
    \begin{feynman}
        \vertex (a) at (0,0);
        \vertex (b) at (-1.5, 1.5);
        \vertex (c) at (-1.5, -1.5);
        \vertex[blob, draw=green, fill=green, minimum size=3mm] (f) at (2.0, 0) {}; 
        \vertex (d) at (3.5, 2.0);
        \vertex (e) at (3.5, -2.0);
        \vertex (g) at (3.5, 1.0);
        \vertex (h) at (3.5, -1.0);
        
        \diagram*  {
            (a) -- [edge label'=$\bar{q}$, fermion] (b),
            (c) -- [edge label=$q$, fermion] (a),
            (a) -- [edge label=$Z / \gamma$, boson] (f),
            (f) -- [edge label=$q$, fermion] (d),
            (g) -- [edge label=$\bar{q}$, fermion] (f),
            (f) -- [edge label=$N$, fermion] (h),
            (e) -- [edge label=$\bar{N}$, fermion] (f),
        };
    \end{feynman}
\end{tikzpicture}
    \caption{}
    \label{fig:QQNNB1}
    \end{subfigure}
    \hfill
    \begin{subfigure}[b]{0.30\textwidth}
        \centering

  \begin{tikzpicture}
  \begin{feynman}
    \vertex[blob, draw=green, fill=green, minimum size=3mm] (a) at (0,0) {}; 
    \vertex (b) at (-1.0, 0.6);
    \vertex (c) at (1.0, 0); 
    \vertex (g) at (2.0, -1.1);
    \vertex (m) at (2.0, -0.1);
    \vertex (n) at (2.0, 1.1);
    \vertex (f) at (0,-1.6);
    \vertex (d) at (1.0, -2.2);
    \vertex (h) at (1.0, -1.2);
    \vertex (e) at (-1.0, -2.2);

    \diagram* {
      (a) -- [edge label=$\bar{q}$, fermion] (b),
      (a) -- [edge label=$N$, fermion] (n),
      (m) -- [edge label'=$\bar{N}$, fermion] (a),
      (g) -- [edge label=$\bar{q}$, fermion] (a),
      (a) -- [edge label'=$Z/\gamma$, boson] (f),
      (f) -- [edge label=$q$, fermion] (d),
      (e) -- [edge label=$q$, fermion] (f),
    };
  \end{feynman}
\end{tikzpicture}

\caption{}
    \label{fig:QQNNB2}
    \end{subfigure}
 \hfill
     \begin{subfigure}[b]{0.30\textwidth}
        \centering
  \begin{tikzpicture}
    \begin{feynman}
        \vertex[blob, draw=green, fill=green, minimum size=3mm] (a) at (0,0) {}; 
        \vertex (b) at (-1.5, 1.5);
        \vertex (c) at (-1.5, -1.5);
        \vertex (f) at (2.0, 1.0);
        \vertex (d) at (3.5, 2.0);
        \vertex (e) at (1.5, -1.5);
        \vertex (g) at (3.5, 1.0);
        \vertex (h) at (1.5, -0.5);
        
        \diagram*  {
            (a) -- [edge label'=$\bar{q}$, fermion] (b),
            (c) -- [edge label=$q$, fermion] (a),
            (a) -- [edge label=$Z / \gamma$, boson] (f),
            (f) -- [edge label=$q$, fermion] (d),
            (g) -- [edge label=$\bar{q}$, fermion] (f),
            (a) -- [edge label=$N$, fermion] (h),
            (e) -- [edge label=$\bar{N}$, fermion] (a),
        };
    \end{feynman}
\end{tikzpicture}

        \caption{}
    \label{fig:QQNNB3}
    \end{subfigure}
  
\captionsetup{justification=raggedright,singlelinecheck=false}     
\caption{Representative Feynman diagrams illustrating the pair production of $N$ within the $\nsmeft$ framework. The red and blue blobs denote D6 operators, while the green blob denotes a D8 operator.}
\label{Fig:LQ_production}
\end{figure}

In this section, we present the cross-section for the process 
$p p \rightarrow j j N N$ mediated by the effective operators 
$\opQ{qN}$, $\opQ{duNe}$, $\opQ{N^2 q^2 B}^{(1)}$, 
and $\opQ{N^2 q^2 B}^{(2)}$ introduced in the previous section. 
Figs.~\ref{fig:DuNe}--\ref{fig:QQNNB3} show the Feynman diagrams corresponding to these operators.

\begin{figure}[h]
    \centering
    \includegraphics[width=0.75\textwidth]{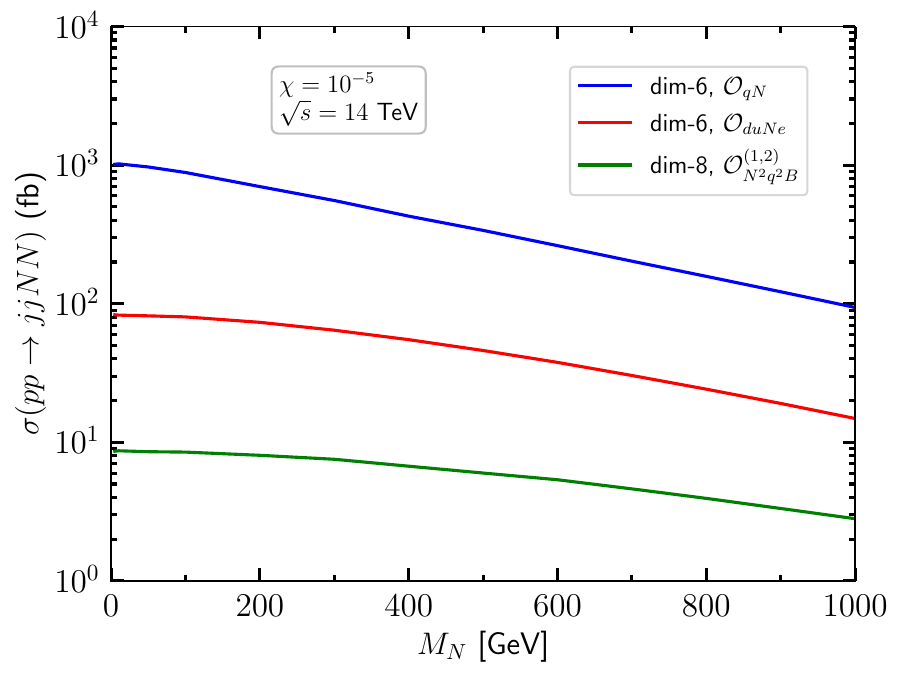}
    
    \caption{Production Cross section of $pp \rightarrow j j N N$ corresponding to different $\nsmeft$ operators.}
    \label{fig:Prod_jjnn}
\end{figure}

Fig.~\ref{fig:Prod_jjnn} illustrates the production cross-sections at the HL-LHC ($\sqrt{s}=14$ TeV) for the $p p \to jjNN$ process as a function of the RHN mass $M_N$, assuming a cutoff scale of $\Lambda = 1$ TeV. The blue, red, and green curves represent the contributions from the D6 operators $\opQ{qN}$ and $\opQ{duNe}$, and the D8 operator $\opQ{N^2 q^2 B}^{(1,2)}$, respectively. As anticipated from phase-space considerations, the cross-sections decrease with increasing $M_N$. Across the entire mass range considered, $\opQ{qN}$ provides the dominant contribution to the production rate, owing to its weaker cutoff suppression at the amplitude level ($\sim 1/\Lambda^2$) compared to the stronger suppression of the other operators. {Note that, although both the contributions of the operators $\opQ{duNe}$ and $\opQ{N^2 q^2 B}^{(1,2)}$ appear  at  $\mathcal{O}({1/\Lambda^4})$, they however significantly differ in orders of magnitude. In the case of $\opQ{N^2 q^2 B}^{(1,2)}$, one of the vertices involves an electroweak gauge coupling, while the contribution from $\opQ{duNe}$ is controlled by  Wilson coefficients of order unity, as evident from the corresponding Feynman diagrams. As a result, the contribution from $\opQ{N^2 q^2 B}^{(1,2)}$ is substantially suppressed compared to that from $\opQ{duNe}$.
}We chose the following Benchmark Points (BP) for the subsequent study.
\begin{itemize}
    \item \textbf{BP1:} \(M_N = 30~\text{GeV},~\Lambda = 1~\text{TeV},~\chi = 10^{-5}\).
    \item \textbf{BP2:} \(M_N = 500~\text{GeV},~\Lambda = 1~\text{TeV},~\chi = 10^{-8}\)
\end{itemize}
In Table~\ref{tab:bench_xsecs}, we present the cross-sections corresponding to these BPs for the process $\sigma_{\rm prod}(pp \rightarrow j j N N)\times {\rm BR}^2(N\to e jj)$.

\begin{table}[h!]
\centering
\begin{tabular}{c|c|c|c}
\hline
\textbf{Benchmark} &
\(\opQ{qN}\) (fb) &
\(\opQ{duNe}\) (fb) &
\(\opQ{N^2 q^2 B}^{(1,2)}\) (fb) \\
\hline
\textbf{BP1} 
& 210.6 & 81.6 & 1.8 \\
\hline
\textbf{BP2} 
& 117.53 & 45.98 & 0.73 \\
\hline
\end{tabular}
\caption{Cross sections 
\(\sigma_{\rm prod}(pp \rightarrow j j N N)\times {\rm BR}^2(N\to \ell jj)\) 
for two benchmark points and different operators.}
\label{tab:bench_xsecs}
\end{table}

\paragraph{Existing Constraints:} Now we discuss the exclusion limits on the Wilson coefficients of the operators under consideration, derived from various collider searches performed at the LHC. The bound on the D6 operator $\opQ{duNe}$ has already been studied conservatively in Ref.~\cite{Mitra:2024ebr}. To constrain the D6 operator $\opQ{qN}$, we choose the mixing parameter $\chi$ and the RHN mass $M_N$ such that the RHN becomes long-lived, leading to displaced decay signatures. Consequently, the relevant experimental probes for our study are the searches targeting displaced final states at the LHC.
\begin{figure}[h]
    \centering
    \begin{minipage}{0.48\textwidth}
        \centering
        \includegraphics[width=\textwidth]{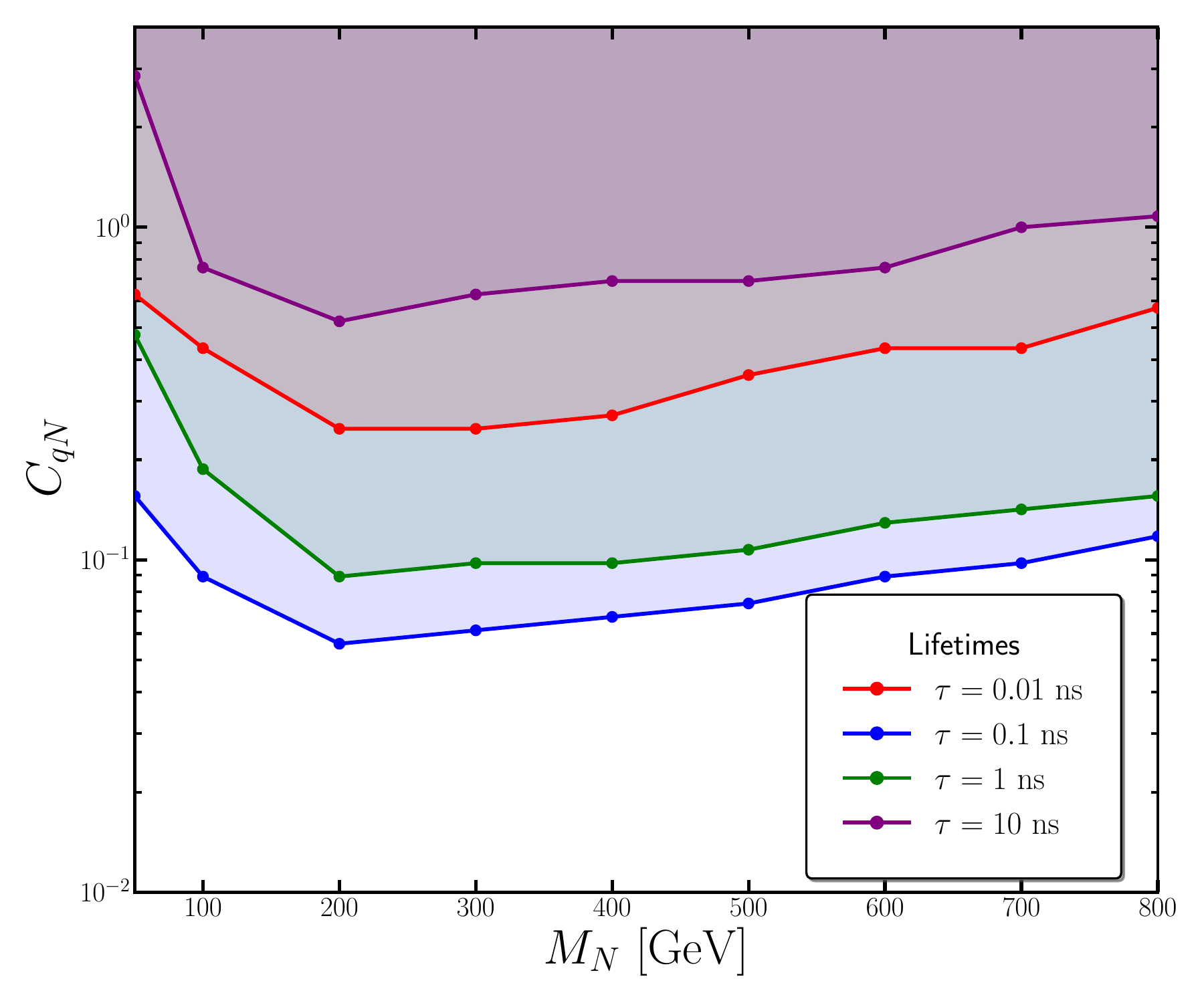}
    \end{minipage}\hfill
    \begin{minipage}{0.48\textwidth}
        \centering
        \includegraphics[width=\textwidth]{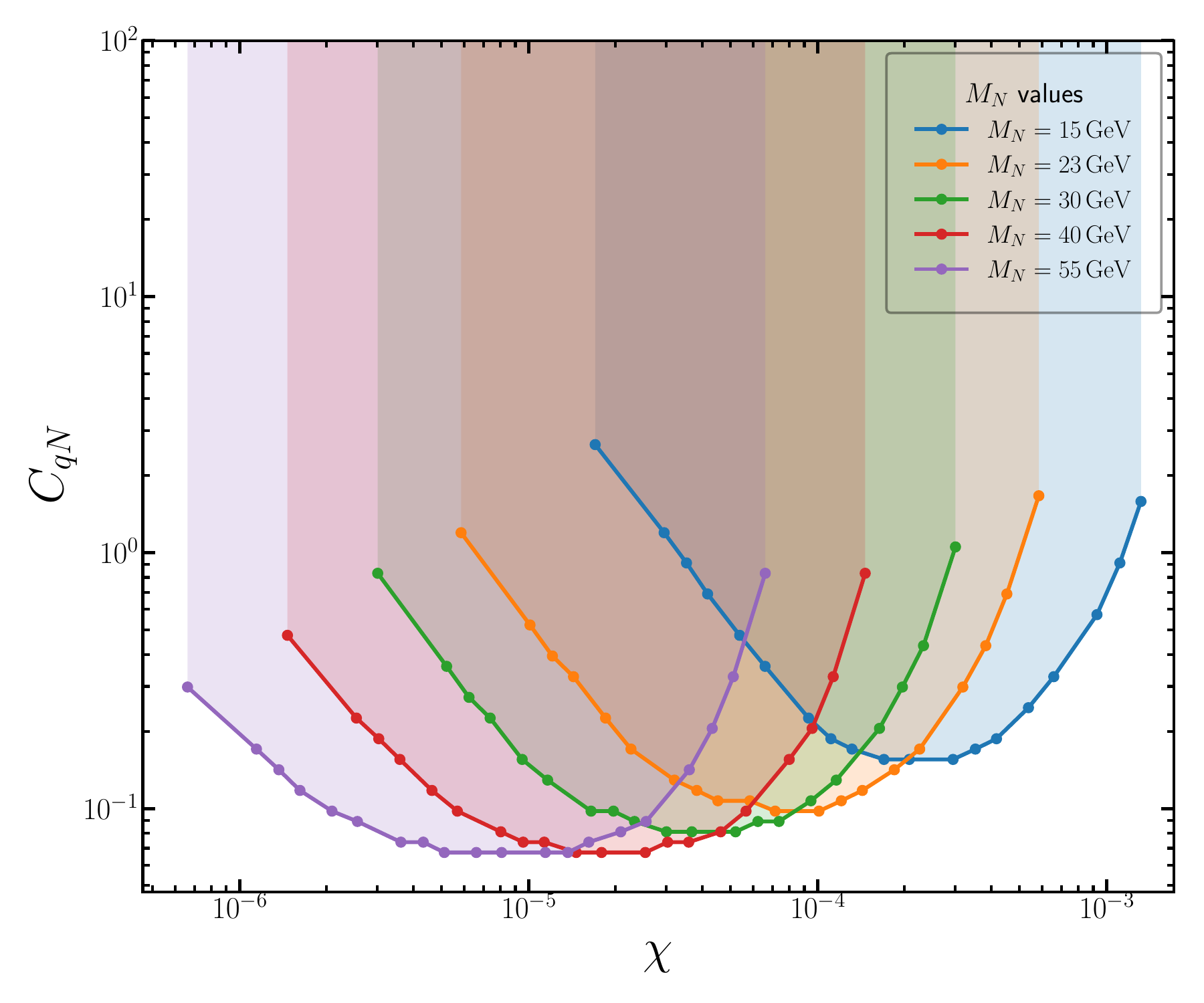}
    \end{minipage}
    \caption{Exclusion limits on the Wilson coefficient \(C_{qN}\) for different benchmark choices of the right-handed neutrino proper lifetime \(\tau\) and mass \(M_N\). The left panel shows \(C_{qN}\) as a function of the RHN mass \(M_N\) for different values of \(\tau\), obtained from the ATLAS search for displaced leptons at \(\sqrt{s} = 13~\mathrm{TeV}\) \cite{ATLAS:2020wjh}. The right panel displays \(C_{qN}\) as a function of the mixing parameter \(\chi\) for different benchmark values of \(M_N\), derived from the CMS search for displaced jets at \(\sqrt{s} = 13.6~\mathrm{TeV}\) \cite{CMS:2024xzb}. The shaded regions represent the experimentally excluded parameter space.
}\label{fig:bound_CQN}
\end{figure}

The left and right panels of Fig.~\ref{fig:bound_CQN} show the bounds on the Wilson coefficient \(C_{qN}\) obtained from the ATLAS search for displaced leptons \cite{ATLAS:2020wjh} at \(\sqrt{s} = 13~\mathrm{TeV}\) and the CMS search for displaced jets \cite{CMS:2024xzb} at \(\sqrt{s} = 13.6~\mathrm{TeV}\), respectively. 

In the left panel, we present the exclusion limits in the $M_N-C_{qN}$ plane for different values of the proper lifetime $(\tau)$, expressed in nanoseconds. The red, blue, green, and violet curves correspond to the maximum allowed values of $C_{qN}$ for $\tau = 0.01, 0.1, 1, \text{ and } 10$~ns, respectively. To derive these bounds, we consider the process
\begin{align*}   
pp \to NN \; (\text{via } \opQ{qN})\;, \qquad N \to e \ell \nu_\ell \; (\text{via }\; \chi)\;,
\end{align*}
and compare the resulting production cross section with the experimental upper limits on the cross section in the SR-ee signal region for the corresponding values of $M_N \equiv M_{\tilde{e}}$ and $\tau$.\footnote{In our setup, the proper lifetime $\tau$ of the RHN is solely determined by the mixing parameter $\chi$ for a given value of $M_N$.} Since the target signal topology in our analysis differs from that considered in the experimental study, the resulting constraints should be regarded as conservative.

In the right panel, we display the exclusion limits in the $\chi-C_{qN}$ plane for several benchmark choices of the RHN mass \(M_N\). The blue, orange, green, red, and violet curves correspond to the maximum allowed values of $C_{qN}$ as functions of $\chi$ for $M_N = 15, 23, 30, 40, \text{ and } 55~\mathrm{GeV}$, respectively. These bounds are obtained by considering the process
\begin{align*}   
pp \to NN \; (\text{via } \opQ{qN}), \qquad N \to e j j \; (\text{via } \chi),
\end{align*}
and comparing the resulting production cross section with the experimental upper limits for the corresponding values of \(M_N \equiv M_{S}\) and \(\chi\).\footnote{In the experimental analysis, several proper decay lengths of \(S \equiv N\) are reported in millimetres. In our framework, these are translated into corresponding values of the mixing parameter \(\chi\), since the decay length is fully determined by \(\chi\).} As in the case of the left panel, the resulting exclusion limits should be interpreted as conservative, owing to the differences between our signal topology and that assumed in the experimental study. {In particular, the experimental search considered in Ref.~\cite{CMS:2024xzb} focuses on final states with displaced hadronic jets arising from Higgs-mediated pair production of long-lived particles, whereas our signal additionally contains a displaced di-electron pair, with the long-lived particle produced via a contact interaction. Due to these differences, the kinematics of final state particles and acceptance of the signal is expected to be somewhat different compared to \cite{CMS:2024xzb}, which will lead to a difference in sensitivity. 
We, however, do not pursue a recast analysis; rather, we present  a conservative estimate. Contrary to the dim-6 operator, the dim-8 operator gives rise to more exotic final states, such as displaced dileptons accompanied by multijets, displaced four-lepton signatures with missing energy, and dijet topologies, which remain unconstrained by existing LHC searches.}

Although the selected benchmark points are not allowed by this very conservative bound, even if this bound is imposed, the discrimination of the $\opQ{qN}$ (D8) contribution becomes clearer, since the bound suppresses only the $\opQ{qN}$ contribution. For a representative scenario, we set $C_{qN}=1$ for both benchmark points, as discussed above.

\subsection{Search Strategy at HL-LHC} \label{subsec:search}

The following analysis employs a BDT framework to distinguish the D8 operator contributions to the $pp \to jj\,N N$ process from two distinct D6 scenarios: (i) the contribution of $\opQ{qN}$, and (ii) the contribution of $\opQ{duNe}$. We implemented the operators of interest using \texttt{FeynRules}~\cite{Alloul:2013bka} to generate the corresponding Universal FeynRules Output (UFO) model files. Signal events were simulated at leading order (LO) using the \texttt{MadGraph5\_aMC@NLO-v3.5.3}~\cite{Alwall:2014hca} event generator, with subsequent parton showering and hadronisation performed via \texttt{Pythia8}~\cite{Sjostrand:2006za}. Notably, SM backgrounds are omitted from this study. For our chosen benchmark parameters (WCs, $\chi$, and $M_N$), the RHN decay is significantly displaced \footnote{{In contrast, as discussed in a later section, the decay mediated by the operator $\mathcal{O}_{duNe}$ leads to prompt signatures. However, since the signal considered here arises from the D8 contribution, its decay is displaced. Consequently, the contribution from SM backgrounds can be safely neglected.}}. This characteristic signature results in a negligible SM background, justifying its exclusion when assessing the discovery potential of these $\nsmeft$ operators.

\begin{figure*}[h]
\centering
\captionsetup[subfigure]{labelformat=empty}
\subfloat[\quad\quad\quad(a)]{\includegraphics[width=0.5\textwidth,height=0.38\textwidth]{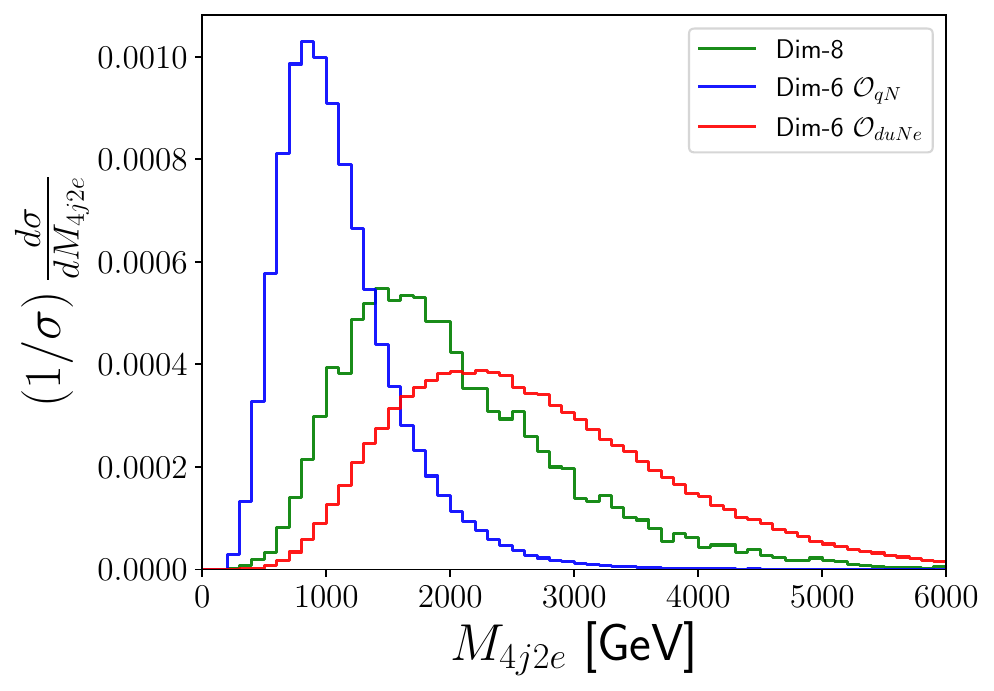}\label{fig:M_4j2l_31July}}
\subfloat[\quad\quad\quad(b)]{\includegraphics[width=0.5\textwidth,height=0.38\textwidth]{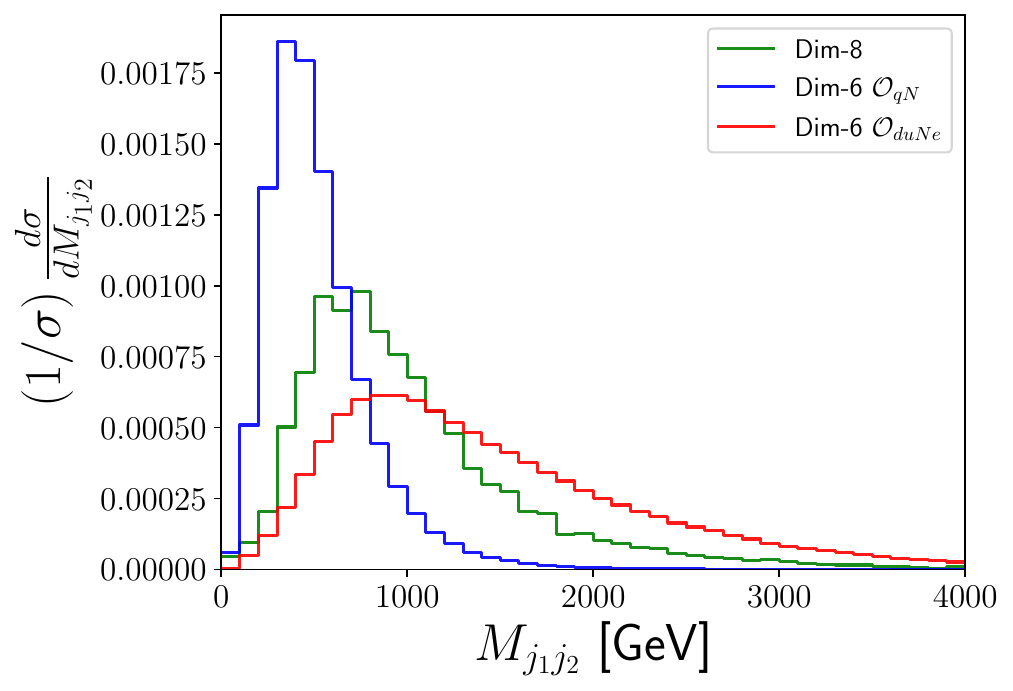}\label{fig:M_j1j2_31July}} \\
\subfloat[\quad\quad\quad(c)]{\includegraphics[width=0.5\textwidth,height=0.38\textwidth]{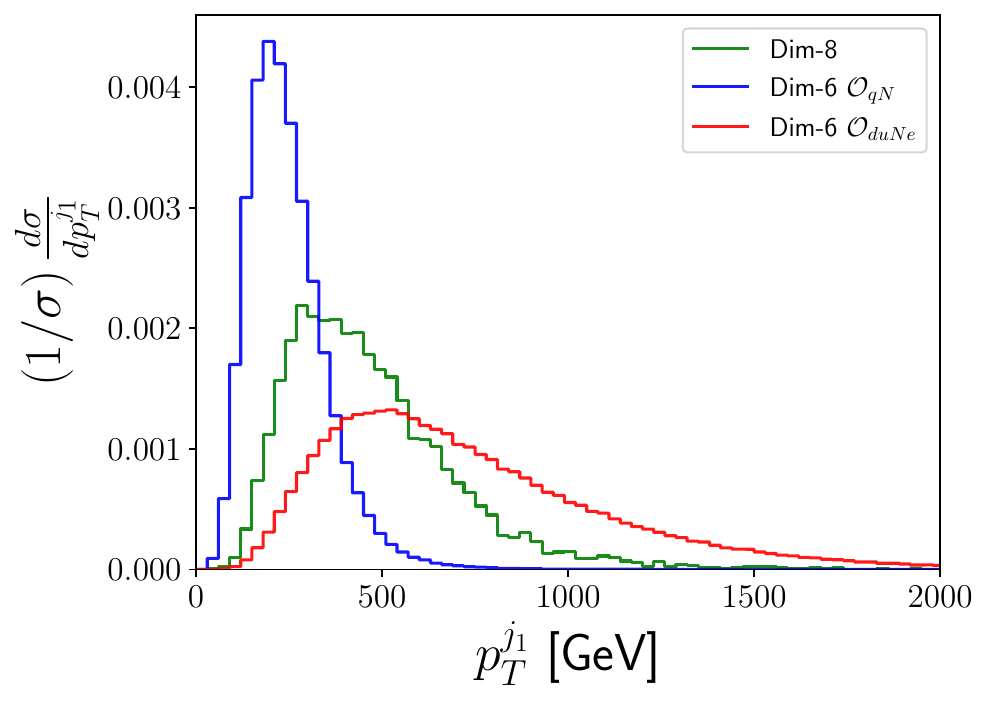}\label{fig:PT_j1_31July}}
\subfloat[\quad\quad\quad(d)]{\includegraphics[width=0.5\textwidth,height=0.38\textwidth]{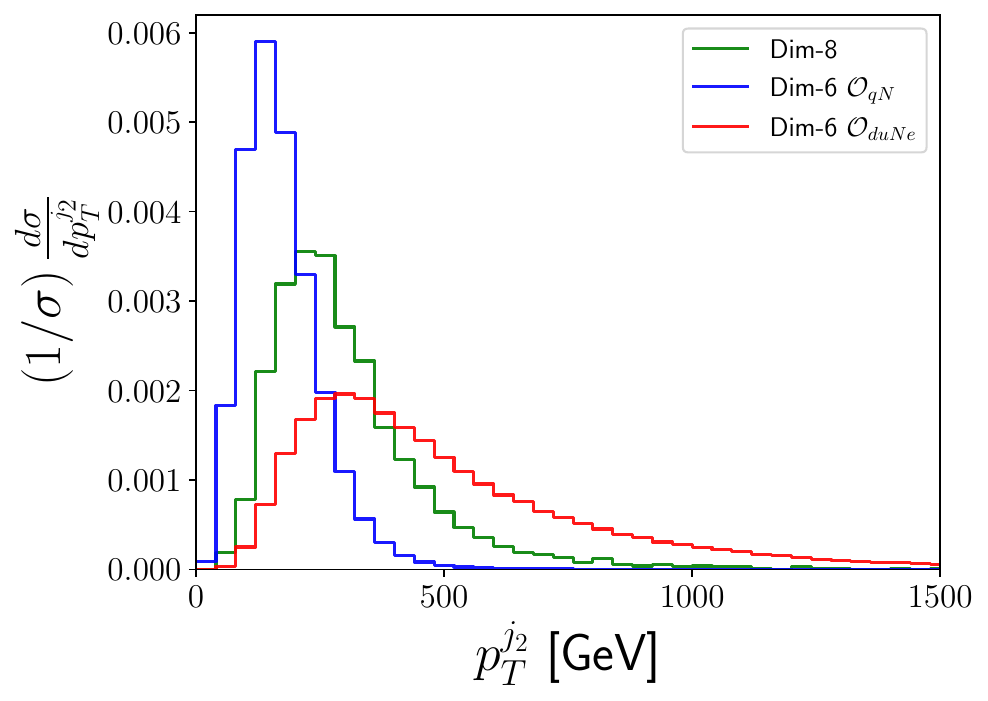}\label{fig:PT_j2_31July}} \\
\subfloat[\quad\quad\quad(e)]{\includegraphics[width=0.33\textwidth,height=0.27\textwidth]{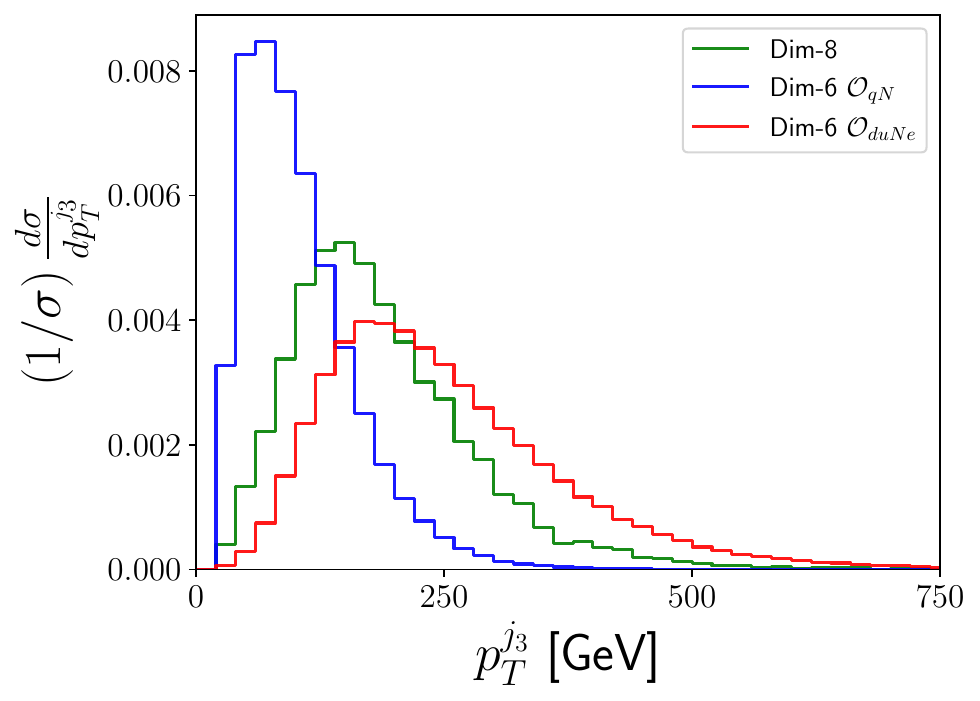}\label{fig:PT_j3_31July}}
\subfloat[\quad\quad\quad(f)]{\includegraphics[width=0.33\textwidth,height=0.27\textwidth]{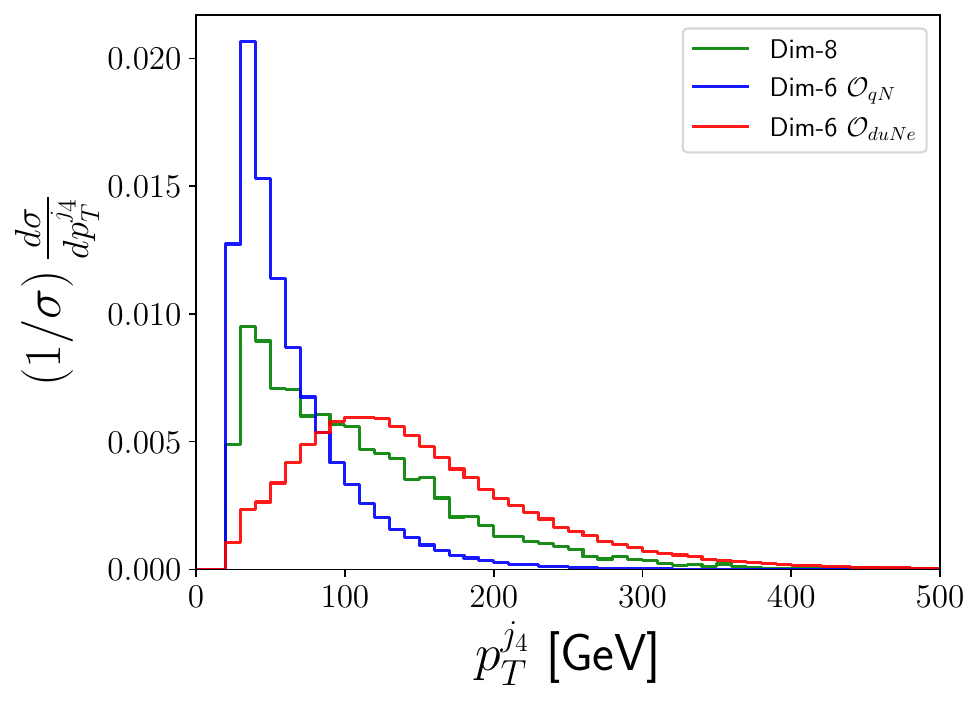}\label{fig:PT_j4_31July}}
\subfloat[\quad\quad\quad(g)]{\includegraphics[width=0.33\textwidth,height=0.27\textwidth]{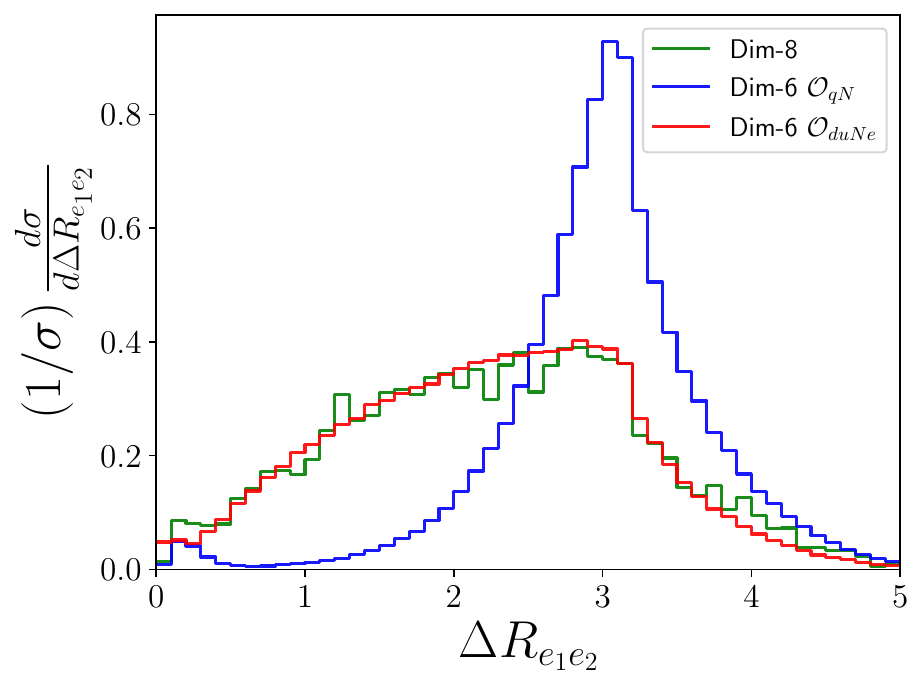}\label{fig:DR_l1l2_31July}} \\
\caption{ The figure showcases kinematic variables for $C_{qN} = 1$, $C_{duNe} = 1$, and $C_{N^2 q^2 B} = 1$ with \textbf{BP1} in $\sqrt{s} = 14$ TeV HL-LHC.
}
\label{fig:kinematicdistribution_dim8_6_compare}
\end{figure*}

\paragraph{Kinematic Variable Selection:} To discriminate the D8 topology from the two D6 cases, we selected 16 kinematic variables as inputs for the multivariate analysis. These are categorised as follows:
\begin{itemize}
    \item Invariant Masses: The total system mass $M_{4j2e}$, the mass of the leading jet pair $M_{j_1 j_2}$, and the single-cluster mass $M_{e j j}$.
    \item Transverse Momenta: The $p_T$ of the four leading jets ($p_{T}^{j_{1–4}}$), the leading and subleading electrons ($p_{T}^{e_{1,2}}$), and the missing transverse energy ($\slashed{E}_T$).
    \item Angular Separations: The $\Delta R$ between jet pairs ($\Delta R_{j_1 j_2}$), electron pairs ($\Delta R_{e_1 e_2}$), and various electron-jet combinations ($\Delta R_{e_i j_k}$).
\end{itemize}
These variables characterise the distinct energy scales and angular correlations inherent to the different operator dimensions. Fig.~\ref{fig:kinematicdistribution_dim8_6_compare} displays the kinematic distributions for the variables most critical to the discrimination process at the 14 TeV HL-LHC. Focusing on benchmark point \textbf{BP1}, we compare the D8 operator (green) against the D6 operators $\opQ{qN}$ (blue) and $\opQ{duNe}$ (red). The primary discriminants—namely $M_{2e4j}$, $M_{j_1 j_2}$, the jet $p_T$ spectrum, and $\Delta R_{e_1 e_2}$—clearly illustrate the harder spectra and distinct topological features associated with the higher-dimensional operator contributions.

After performing the BDT analysis for both benchmarks, we obtain significance greater than $5 \sigma$, from which the D8 contribution can be classified in both D6 scenarios.

\section{Results} \label{sec:results}

\begin{figure}[h]
\centering
\includegraphics[width=0.6\textwidth]{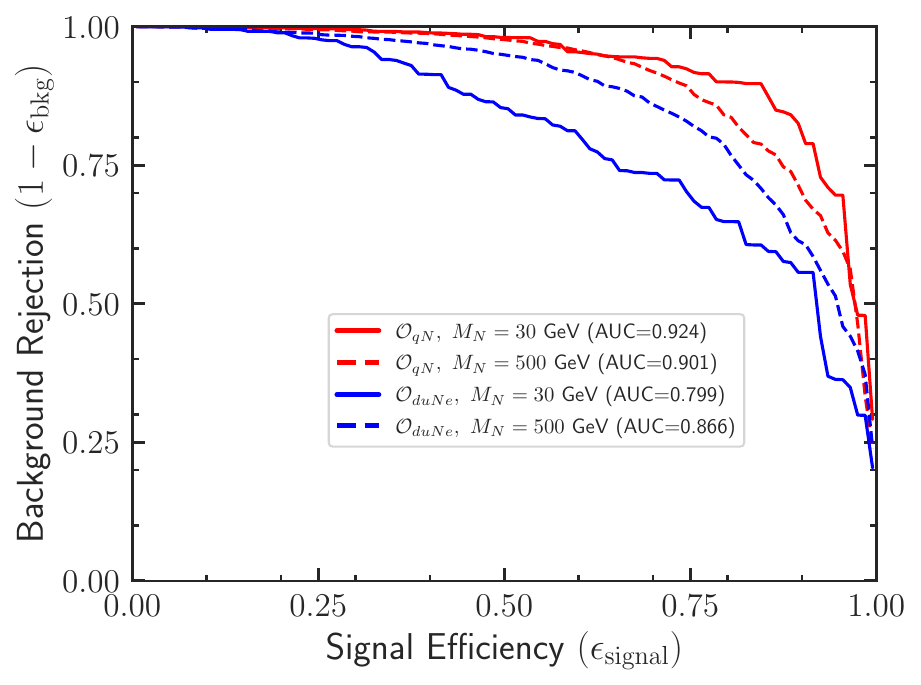}
\caption{
The figure showcases the ROC Curve for the BDT classifier for two cases : (i) and (ii)  for \textbf{BP1} and \textbf{BP2} at $\sqrt{s} = 14$ TeV HL-LHC. In the x-axis, Signal refers to the dim-8 contribution, and in the y-axis, Background implies the corresponding dim-6 operator contributions.}
\label{fig:ROC}
\end{figure}

The performance of the BDT analysis for the two classification scenarios—case~(i) (D8 vs. $\opQ{qN}$) and case (ii) (D8 vs. $\opQ{duNe}$)—is evaluated using the ROC curves presented in Fig.~\ref{fig:ROC}.
In this plot, the red and blue curves represent case (i) and case (ii), respectively, while the solid and dashed line styles distinguish between benchmark points \textbf{BP1} and \textbf{BP2}. The high AUC values across these scenarios demonstrate the robust discriminatory power of the chosen kinematic features. Detailed performance metrics and numerical results for these benchmark points are summarised in Table~\ref{tab:bench_results}.For \textbf{BP1}, case (i) yields a significance of \( 17.28\,\sigma \), while case (ii) gives \( 17\,\sigma \) for the classification.
\begin{table}[h!]
\centering
\setlength{\tabcolsep}{5pt}  
\renewcommand{\arraystretch}{1.2} 

\begin{tabular}{c|c|c|c|c}
\hline
\textbf{Benchmark} &
AUC case (i) &
AUC case (ii) &
Sig. case (i) &
Sig. case (ii) \\
\hline
\textbf{BP1} & 0.924 & 0.799 & \(17.28\,\sigma\) & \(17\,\sigma\) \\
\hline

\textbf{BP2} & 0.901 & 0.866 & \(10.68\,\sigma\) & \(13.20\,\sigma\) \\
\hline

\end{tabular}

\caption{Classification performance (AUC and significance) 
for two benchmark points comparing the D8 operator $\opQ{N^2 q^2 B}$ 
against the D6 operators $\opQ{qN}$ and $\opQ{duNe}$.}
\label{tab:bench_results}
\end{table}
Although case (ii) exhibits a lower AUC compared to case (i), its final classification sensitivity remains comparable. This is primarily because the D6 contribution in case (ii) has a lower production-level cross section, which enhances the relative impact of the D8 contribution in the classification. A similar effect is observed for \textbf{BP2} as well.

It is important to emphasise that for case (ii), our classification strategy intentionally omits information regarding the RHN decay length or associated displaced-vertex observables. The inclusion of such signatures would markedly enhance the discriminatory power, given that the $\opQ{duNe}$ operator mediates a significantly shorter decay length compared to the D8 operators. Nevertheless, even in the absence of this distinctive kinematic handle, the current analysis demonstrates robust statistical significance for the selected benchmark points.


\section{Conclusion} \label{sec:conclusion}

In this work, we have investigated the effects of dimension-eight operators in the $\nsmeft$ framework, with a particular focus on their collider phenomenology and their distinction from leading dimension-six contributions. Using the Hilbert series formalism, we derived the complete and independent basis of D8 operators involving right-handed neutrinos and verified that the resulting operator counting is in full agreement with the literature. This provides a consistent and well-defined foundation for studying higher-dimensional effects in scenarios where right-handed neutrinos are kinematically accessible at collider energies.

Building on this theoretical construction, we performed a phenomenological analysis of the representative D8 operator $\opQ{N^{2}q^{2}B}^{(1,2)}$ at the HL-LHC. We considered pair production of right-handed neutrinos in association with jets, followed by their decays into electron-jet final states, yielding signatures with at least four jets and two electrons. The analysis was performed in benchmark scenarios where the right-handed neutrinos are long-lived, leading to displaced decay vertices and effectively negligible Standard Model backgrounds. This setup allowed us to isolate genuinely new event topologies that do not arise at dimension six.

A central objective of this study was to determine whether the effects of the selected D8 operator can be experimentally distinguished from those induced by leading D6 operators that generate similar final states. To address this question in a controlled manner, we treated the D6 operators $\opQ{qN}$ and $\opQ{duNe}$ independently as competing hypotheses. We employed a Boosted Decision Tree multivariate analysis based on a set of carefully chosen kinematic observables sensitive to the different energy scaling and topology of the underlying interactions. In particular, the invariant masses $M_{4j2e}$ and $M_{j_1 j_2}$, the transverse momentum spectra of the leading jets, and the angular separation $\Delta R_{e_1 e_2}$ were identified as the most powerful discriminating variables.

For two representative benchmark points, spanning light and heavy right-handed neutrino masses, we demonstrated that the D8 contribution can be robustly separated from each of the considered D6 scenarios. In both cases, the classification achieves statistical significances well above $5\sigma$ at the HL-LHC. Notably, this discrimination is achieved using only prompt kinematic information, without exploiting additional observables related to decay lengths or displaced-vertex reconstruction. The inclusion of such information would be expected to further enhance the sensitivity to higher-dimensional operators.

Our results show that dimension-eight operators in the $\nsmeft$ can lead to experimentally distinguishable signatures at the HL-LHC and that multivariate analysis techniques provide a powerful tool for resolving the underlying EFT structure. This highlights the importance of extending SMEFT-based collider studies beyond dimension six when interpreting future data in scenarios with light right-handed neutrinos. Future work may explore additional D8 operator classes, incorporate a broader set of dimension-six backgrounds, and investigate more advanced machine-learning strategies to further strengthen the discovery and interpretation potential of $\nsmeft$ searches.

\acknowledgments
S.U.R. and M.S. are supported by the STFC under
grant ST/P001246/1. M.M  acknowledges IPPP DIVA Award research programme for support.

\appendix
\section{Operator Constrution with Hilbert Series} \label{app:op}
The Hilbert Series framework is thoroughly detailed in the following refs~\cite{Banerjee:2020bym,Banerjee:2020jun,Graf:2022rco,Hanany:2010vu,Henning:2015alf,Henning:2017fpj,Lehman:2015via,Lehman:2015coa,Sun:2022aag,Grinstein:2024iyf,Grinstein:2024jqt}.  To determine the invariant, independent set of operator structures with HS, we must first define the characters of each field according to their transformation properties, as summarised in Table~\ref{tab:vSMEFT-field-content}. The invariance is then enforced by carrying out the integration over the group volume, i.e., the Haar measure. This step leverages the orthogonality property of the characters to project out only the invariant structures. Schematically, HS is obtained by
\begin{align}
    \mathcal{H} = \int d \mu \; \Bigg(\prod_i PE_i[\phi_i,R_i]\Bigg)\;,
\end{align}
where, $d\mu$ is the Haar measure for all the groups including the internal symmetry group $(SU(3)_c\times SU(2)_w\times U(1)_y)$ and spacetime symmetry group $(SO(3,1)\sim SU(2)_L\times SU(2)_R)$ and $PE[\phi, R]$ denotes the plethystics for each field $\phi_i$. It is a function of the combination of the characters $(R)$ for each field.  Since performing the integration is still not feasible, we use the following Mathematica package~\cite{Alonso:2024usj,Banerjee:2020bym} to compute the HS output. In the following, we will discuss the remaining operators that complete the basis. 

\subsection{Operators and Counting}

We organise the effective operators hierarchically for clarity. The broadest grouping is the operator class, which aggregates fields with the same spin and number of derivatives. Within each operator class, operators are further separated into different 'types', where operators sharing the same field content are grouped together. This organisation is reflected across Tables~\ref{tab:dim8-ops-psi4x-1} through \ref{tab:dim8-ops-4}, where we present these different operator 'types'. 
\begin{table}[h]
    \centering
    \renewcommand{\arraystretch}{2.1}
    {\begin{tabular}{c|c}
    \multicolumn{2}{c}{$\mathbf{\Psi^4 X } \;\;(+\hc)$}\\
        \hline
        \hline
        \opq{dNq^2B}{1}&
        $\blv \epsilon_{\alpha \beta \gamma} \epsilon_{jk}(q^{j \alpha} C \sigma^{\mu \nu} q^{k \beta})(d^{\gamma} C N) B_{\mu \nu}  $ \\
        
        \opq{dNq^2B}{2} & 
        $\blv \epsilon_{\alpha \beta \gamma} \epsilon_{jk}(q^{j \alpha} C  q^{k \beta})(d^{\gamma} C \sigma^{\mu \nu} N) B_{\mu \nu} $ \\
        
        \opq{dNq^2G}{1} & 
        $ \blv{(T^{A})}_{\gamma}^{\delta} \epsilon_{\delta \alpha \beta} \epsilon_{jk} (q^{j \alpha} C \sigma^{\mu \nu} q^{k \beta})(d^{\gamma} C N) G_{\mu \nu}^{A}  $ \\
        
        \opq{dNq^2G}{2} &
        $\blv {(T^{A})}_{(\alpha}^{\delta} \epsilon_{\beta ) \gamma \delta} \epsilon_{jk} (q^{j \alpha} C  q^{k \beta})(d^{\gamma} C \sigma^{\mu \nu} N) G_{\mu \nu}^{A} $ \\
        
        $\opQ{d^2NuB} $& 
        $\blv  \epsilon_{\alpha \beta \gamma}(u^{\alpha} C \sigma^{\mu \nu} d^{\beta})(d^{\gamma}C N)B_{\mu \nu} $ \\
        
        $\opQ{d^2NuG}$& 
        $ \blv {(T^{A})}_{\gamma}^{\delta} \epsilon_{\delta \alpha \beta} (u^{ \alpha} C \sigma^{\mu \nu} d^{ \beta})(d^{\gamma} C N) G_{\mu \nu}^{A} $ \\ 
        
        \opq{dNq^2W}{1}& 
        $ \blv \epsilon_{\alpha \beta \gamma} {(\epsilon \tau^{I})}_{jk} (q^{j \alpha} C \sigma^{\mu \nu} q^{k \beta})(d^{\gamma} C N) W_{\mu \nu}^{I}  $ \\
        
        \opq{dNq^2W}{2}& 
        $ \blv \epsilon_{\alpha \beta \gamma} {(\epsilon \tau^{I})}_{jk} (q^{j \alpha} C  q^{k \beta})(d^{\gamma} C \sigma^{\mu \nu} N_R) W_{\mu \nu}^{I} $ \\

        $\opQ{N^4B}$& 
        $ \blv (N C \sigma^{\mu \nu} N)(N C  N) B_{\mu \nu} $ \\
        \hline
        \end{tabular}}
	\caption{Continued from Table~\ref{tab:dim8-ops-psi4x-1}. $B,L$-violating operators of $\mathbf{\Psi^4 X }$ class }
	\label{tab:dim8-ops-psi4x-2}
\end{table}

\begin{table}[h]
    \centering
    \renewcommand{\arraystretch}{2.2}
    {\small\begin{tabular}{c|c|c|c}
        \multicolumn{2}{c|}{$\mathbf{\Psi^2 H^5 +\hc}$} & \multicolumn{2}{c}{$\mathbf{\Psi^2H^4 D} $}\\
        \hline
        \hline
        \multirow{2}{*}{$\opQ{H^5lN}$} & \multirow{2}{*}{
        $ (H^{\dagger} H)^2(\bar{l} N \tilde{H}) $ } &
        $\opQ{H^3eND}$ &
        $ (\bar{e} \gamma^{\mu} N) (H^{\dagger} \overleftrightarrow{D_{\mu}} \tilde{H})(H^{\dagger} H)\; (+\hc) $ \\
        & &
        $\opQ{H^3N^2D}{}$& 
        $ (\bar{N} \gamma^{\mu} N) (H^{\dagger} \overleftrightarrow{D_{\mu}} H)(H^{\dagger} H) $ \\
        \hline
    \end{tabular}}
    \caption{Operators in the \(\mathbf{\Psi^2 H^5 }\) class (left panel) and the \(\mathbf{\Psi^2 H^4 D}\) class (right panel). $(+\hc)$ denotes the hermitian conjugate of that operator.}
    \label{tab:dim8-ops-3a}
\end{table}

\begin{table}[h]
    \centering
    \renewcommand{\arraystretch}{2.2}
    {\small\begin{tabular}{c|c|c|c}
        \multicolumn{4}{c}{$\mathbf{\Psi^2 H^3X + \hc}$}\\
        \hline
        \hline
        $\opQ{H^3lNB}$ &
        $ (\bar{l}^j \sigma^{\mu \nu} N) \tilde{H}_j (H^{\dagger} H) B_{\mu \nu} $ &
        $\opQ{H^3lNW}$ &
        $  (\bar{l}^j \sigma^{\mu \nu} N) (\tau^{I} \tilde{H})_j (H^{\dagger} H) W_{\mu \nu}^{I} $
        \\
        \hline
    \end{tabular}}
    \caption{Operators in $\mathbf{\Psi^2 H^3 X }$ class. $(+\hc)$ denotes the hermitian conjugate of that operator. }
    \label{tab:dim8-ops-3b}
\end{table}

\begin{table}[h]
    \centering
    \renewcommand{\arraystretch}{2.2}
    {\small\begin{tabular}{c|c|c|c}
        \multicolumn{4}{c}{} \\
        \multicolumn{4}{c}{$\mathbf{\Psi^2 H^3 D^2 + \hc}$}\\
        \hline
        \hline
        $\opQ{H^3lND^2}$ &
        \multicolumn{3}{c}{
        $ (D_{\mu} H^{\dagger} D^{\mu} H)(\bar{l} N \tilde{H}) $ } \\
        \hline
        \end{tabular}}
         \caption{Operators in $\mathbf{\Psi^2 H^3 D^2 }$ class. $(+\hc)$ denotes the hermitian conjugate. }
	\label{tab:dim8-ops-3}
\end{table}

\begin{table}[h]
    \centering
    \renewcommand{\arraystretch}{2.2}
    {\small\begin{tabular}{c|c|c|c}
        \multicolumn{4}{c}{} \\
        \multicolumn{4}{c}{$\mathbf{\Psi^2 H^2 X D \;\;(+\hc)}$}\\
        \hline
        \hline
        $\opQ{H^2eNBD}^{(1)}$ & 
        $ (\bar{N} \gamma^{\nu} e)(\tilde{H}^{\dagger} \overleftrightarrow{D^{\mu}} H) B_{\mu \nu}$ & 
        $\opQ{H^2eNBD}^{(2)}$&  
        $ (\bar{N} \gamma^{\nu} e)(\tilde{H}^{\dagger} \overleftrightarrow{D^{\mu}} H) \tilde{B}_{\mu \nu} $\\
        $\opQ{H^2eNWD}^{(1)}$&
        $ (\bar{N} \gamma^{\nu} e)(\tilde{H}^{\dagger} \overleftrightarrow{D^{I \mu}} H) W_{\mu \nu}^{I}  $ &
        $\opQ{H^2eNWD}^{(2)}$ &
        $ (\bar{N} \gamma^{\nu} e)(\tilde{H}^{\dagger} \overleftrightarrow{D^{I \mu}} H) \tilde{W}_{\mu \nu}^{I} $ \\
        $\opQ{H^2N^2BD}^{(1)}$&
        $ (\bar{N} \gamma^{\nu} N) D^{\mu}(H^{\dagger} H) B_{\mu \nu}  $ &
        $\opQ{H^2N^2BD}^{(2)}$&
        $ (\bar{N} \gamma^{\nu} N) D^{\mu}(H^{\dagger} H) \tilde{B}_{\mu \nu}  $\\
        \hline
	\end{tabular}}
	\caption{Operators in $\mathbf{\Psi^2 H^2 X D }$ class. $(+\hc)$ denotes the hermitian conjugate of that operator. }
	\label{tab:dim8-ops-1}
\end{table}

\begin{table}[h]
    \centering
    \renewcommand{\arraystretch}{2.2}
    {\small\begin{tabular}{c|c|c|c}
        \multicolumn{4}{c}{$\mathbf{\Psi^4 H^2}$}\\
        \hline
        \hline
        $\opQ{H^2e^2N^2}$&
        $(\bar{e} \gamma^{\mu} e) (\bar{N} \gamma^{\mu} N) (H^{\dagger} H)$ &
        $\opQ{H^2l^2N^2}$ &
        $(\bar{l} \gamma^{\mu} l)(\bar{N} \gamma_{\mu} N)(H^{\dagger} H) $ \\
        
        $\opQ{H^2d^2N^2}$&
        $(\bar{d} \gamma^{\mu} d) (\bar{N} \gamma_{\mu} N)(H^{\dagger} H)  $ &
        $\opQ{H^2q^2N^2}$ &
        $(\bar{q} \gamma^{\mu} q)(\bar{N} \gamma_{\mu} N)(H^{\dagger} H) $ \\
        
        $\opQ{H^2u^2N^2}$&
        $ (\bar{u} \gamma^{\mu} u)(\bar{N} \gamma_{\mu} N)(H^{\dagger} H) $ &
        $\opQ{H^2N^4}$&
        $  (\bar{N} \gamma^{\mu} N)(\bar{N} \gamma_{\mu} N)(H^{\dagger} H)$ \\
        
        $\opQ{H^2el^2N}^{(1)}$ & 
        $ (\bar{l} N \tilde{H})(H^{\dagger} \bar{e} l) \;\;(+\hc)  $ &
        $\opQ{H^2el^2N}^{(2)}$ &
        $ (\bar{l}^{j} N) \epsilon_{jk} (\bar{l}^{k} e)(H^{\dagger} H) \;\;(+\hc)$ \\
        
        \opq{H^2dlNq}{1} &
        $ (\bar{l}^{j} N) \epsilon_{jk} (\bar{q}^{k} d)(H^{\dagger} H) \;\;(+\hc)$ &
         \opq{H^2dlNq}{2}&
        $  (\bar{l}^{j} N)(\tau^{I} \epsilon)_{jk} (\bar{q}^{k} d)(H^{\dagger} \tau^{I} H) \;\;(+\hc)$ \\
        
        \opq{H^2lNqu}{1}&
        $  (\bar{l}^j N)(\bar{u} q_j)(H^{\dagger} H) \;\;(+\hc)$ &
        \opq{H^2lNqu}{2}&
        $ (\bar{l}^j N) (\tau^{I})_{jk} (\bar{u} q^k)(H^{\dagger} \tau^{I} H) \;\; (+\hc) $ \\
        
        $\opQ{H^2deNu}$& 
        $ (\bar{u} \gamma^{\mu} d)(\bar{e} \gamma_{\mu} N)(H^{\dagger} H) \;\;(+ \hc) $ &
        $\opQ{H^2eNq^2}$& 
        $ (\bar{q}^\alpha N \tilde{H})(H^{\dagger} \bar{e} q_\alpha) \;\;(+\hc) $ \\
        
        \opq{H^2l^2N^2}{1}&
        $ (\bar{l} N \tilde{H})(\bar{l} N \tilde{H}) \;\;(+\hc)$ &
        & \\
        \hline
        \hline
        $\opQ{H^2dNq^2}$& 
         \blvmc{${\epsilon}_{\alpha \beta \gamma} {\epsilon}_{jk} (d^{\alpha} C {N})(q^{j \beta } C q^{k \gamma }) (H^{\dagger} H) \;(+\hc) $} \\
        
        $\opQ{H^2d^2Nu}$ & 
        \blvmc{$ {\epsilon}_{\alpha \beta \gamma} (d^{\alpha} C {d}^{\beta})(u^{\gamma } C N) (H^{\dagger} H) \;(+\hc)  $} \\
        
        $\opQ{H^2Nq^2u}$& 
        \blvmc{  $ (\epsilon_{\alpha \beta \gamma} \epsilon_{jk} \epsilon_{mn}) (q^{\alpha j} C q^{\beta k}) (u^{\gamma} C N_R)({\tilde{H}}^{m}{\tilde{H}}^{n}) \; (+\hc) $} \\
        
        $\opQ{H^2l^2N^2}$& 
        \blvmc{  $ (\epsilon_{jk} \epsilon_{mn})(l^{j} C l^{k}) (N C N_R)({\tilde{H}}^{m}{\tilde{H}}^{n}) \;(+\hc) $} \\
        
        $\opQ{H^2N^4}$& 
        \blvmc{  $(N C N) (N C N)(H^{\dagger} H)\; (+\hc)  $} \\
        \hline
	\end{tabular}}
	\caption{Operators in $\mathbf{\Psi^4 H^2 }$ class. $(+\hc)$ denotes the hermitian conjugate of that operator. The operators listed in the lower section of the table violate Baryon and/or Lepton numbers.}
	\label{tab:dim8-ops-psi4h2}
\end{table}

\begin{table}[h]
    \centering
    \renewcommand{\arraystretch}{2.2}
    {\begin{tabular}{c|c|c|c}
\multicolumn{4}{c}{$\mathbf{\Psi^4 D^2}$}\\
        \hline
        \hline
        $\opQ{d^2N^2D^2}$ &
        $ D^{\nu} (\bar{d} \gamma^{\mu} d) D_{\nu} (\bar{N} \gamma_{\mu} N) $ &
        $\opQ{e^2N^2D^2}$ &
        $ D^{\nu} (\bar{e} \gamma^{\mu} e) D_{\nu} (\bar{N} \gamma_{\mu} N) $ \\
        
        $\opQ{l^2N^2D^2} $&
        $  D^{\nu} (\bar{l} \gamma^{\mu} l) D_{\nu} (\bar{N} \gamma_{\mu} N) $ &
        $\opQ{N^2q^2D^2}$ &
        $  D^{\nu} (\bar{q} \gamma^{\mu} q) D_{\nu} (\bar{N} \gamma_{\mu} N) $ \\
        
        $\opQ{N^2u^2D^2}$ &
        $  D^{\nu} (\bar{u} \gamma^{\mu} u) D_{\nu} (\bar{N} \gamma_{\mu} N) $ &
        $\opQ{N^4D^2}$&
        $  D^{\nu} (\bar{N} \gamma^{\mu} N) D_{\nu} (\bar{N} \gamma_{\mu} N) $ \\

        $\opQ{el^2ND^2}$ & 
        $ D_{\mu} (\bar{l^{j}} e) \epsilon_{jk} D^{\mu} (\bar{l^{k}} N) \;\;(+\hc) $ &
        $\opQ{dlqND^2}$ &
        $  D_{\mu} (\bar{l^{j}} N) \epsilon_{jk} D^{\mu} (\bar{q^{k}} d) \;\;(+\hc)$\\
        
        $\opQ{deNuD^2}$ &
        $ D^{\nu} (\bar{e} \gamma^{\mu} N) D_{\nu} (\bar{u} \gamma_{\mu} d) \;\;(+\hc)$ &
        $\opQ{lNquD^2}$ &
        $ D^{\nu} (\bar{l} \gamma^{\mu} q^{\alpha}) D_{\nu} (\bar{u_{R \alpha}} \gamma_{\mu} N) \;\;(+\hc)$ \\
        \hline
        \hline
        $\opQ{Nq^2uD^2}$&
        \blvmc{$ \epsilon_{\alpha \beta \gamma } \epsilon_{jk} (q^{j \alpha} C D_{\mu} q^{k \beta}) D^{\mu} (d^{\gamma} C N)\;\;(+\hc) $ } \\
        
        $\opQ{d^2NuD^2}$ &
        \blvmc{$ \epsilon_{\alpha \beta \gamma}(d^{\alpha} C D_{\mu} d^{\beta} )D^{\mu}(u^{\gamma} C N ) \;(+\hc)$} \\
        $\opQ{N^4D^2}$&
        \blvmc{$  (N C D_{\mu} N) D^{\mu} (N C  N)\;\;(+\hc) $ }\\
        \hline
        \end{tabular}}
	\caption{Operators in $\mathbf{\Psi^4 D^2 }$ class. $(+\hc)$ denotes the hermitian conjugate of that operator. The operators listed in the lower section of the table violate Baryon and/or Lepton numbers. }
	\label{tab:dim8-ops-psi4d2}
\end{table}

\begin{table}[h]
    \centering
    \renewcommand{\arraystretch}{2.2}
    {\small\begin{tabular}{c|c|c|c}
        \multicolumn{2}{c|}{$\mathbf{\Psi^2 H^2 D^3}$} & \multicolumn{2}{c}{$ \mathbf{\Psi^2 X^2 D}$}\\
        \hline
        \hline
        $\opQ{H^2N^2D^3}$&
        $ (\bar{N} \gamma^{\mu} D^{\nu} N )( D_{(\mu} D_{\nu )} H^{\dagger} H) $&
        $\opQ{N^2B^2D}$&
        $ (\bar{N_{R}} \gamma^{\mu} \overleftrightarrow{D^{\nu}} N_{R})B_{\mu \rho} B_{\nu}^{\rho} $ \\
        $\opQ{H^2N^2D^3}$&
        $(\bar{N_{R}} \gamma^{\mu} D^{\nu} N_{R}) ( \tilde{H}^{\dagger} D_{(\mu} D_{\nu )} H)  + \hc $ &
        $\opQ{N^2G^2D}$&
        $ (\bar{N_{R}} \gamma^{\mu} \overleftrightarrow{D^{\nu}} N_{R})G_{\mu \rho}^{A} G_{\nu}^{A \rho} $ \\
        & 
        &
        $\opQ{N^2W^2D}$&
        $  (\bar{N_{R}} \gamma^{\mu} \overleftrightarrow{D^{\nu}} N_{R})W_{\mu \rho}^{I} W_{\nu}^{I \rho} $
        \\
        \hline
        \end{tabular}}
        \caption{Operators in the \(\mathbf{\Psi^2 H^2 D^3}\) class (left panel) and the \(\mathbf{\Psi^2 X^2 D}\) class (right panel). }

    \centering
    \renewcommand{\arraystretch}{2.2}
    {\small\begin{tabular}{c|c|c|c}
        \multicolumn{4}{c}{$ \mathbf{\Psi^2 H X D^2} + \hc$}\\
        \hline
        \hline
        \opq{HlNBD^2}{1}& 
        $  (\bar{l} \sigma^{\mu \nu} D^{\rho} N) (D_{\nu} \tilde{H}) B_{\rho \mu} $ &
        \opq{HlNWD^2}{1} &
        $ (\bar{l} \sigma^{\mu \nu} D^{\rho} N) \tau^{I} (D_{\nu} \tilde{H}) W_{\rho \mu}^{I} $ \\
       
        \opq{HlNBD^2}{2}&
        $ (\bar{l} D^{\rho} N) (D^{\nu} \tilde{H}) \tilde{B}_{\rho \mu} $ &
        \opq{HlNWD^2}{2} &
        $  (\bar{l} D^{\rho} N) \tau^{I} (D^{\nu} \tilde{H}) \tilde{W}_{\rho \mu}^{I} $\\
        \hline
        \end{tabular}}
	\caption{Operators in $\mathbf{\Psi^2 H X D^2 }$ class. $(+\hc)$ denotes the hermitian conjugate of that operator. }
	\label{tab:dim8-ops-3}
\end{table}
\clearpage
\newpage
\begin{table}[h]
    \centering
    \renewcommand{\arraystretch}{2.2}
    {\small\begin{tabular}{c|c|c|c}
        \multicolumn{4}{c}{$\mathbf{\Psi^4 H D \;\;(+\hc)}$}\\
        \hline
        \hline
        $\opQ{Hd^2lND}$ & 
        $ (\bar{d} \gamma^{\mu} d) (\bar{l}^j N) (D_{\mu} \tilde{H})_j $& 
        $\opQ{He^2lND}$&
        $  (\bar{e} \gamma^{\mu} e) (\bar{l}^j D_{\mu} N) \tilde{H}_j $ \\
        
        $\opQ{Hl^3ND}$ &
        $ (\bar{l} \gamma^{\mu} l) (\bar{l}^j N) (D_{\mu} \tilde{H})_j $ &
        $\opQ{HedqND}$ &
        $ (\bar{e} \gamma^{\mu} N) (\bar{q} d) D_{\mu} \tilde{H} $ \\
        
        $\opQ{Hlq^2ND}$ &
        $ (\bar{q} \gamma^{\mu} q) (\bar{l} N) D_{\mu} \tilde{H} $ &
        
        $\opQ{HdlNuD}$ &
        $ (\bar{u} \gamma^{\mu} d) (\bar{l} N) D_{\mu} H $ \\

        $\opQ{HeNquD}$ &
        $ (\bar{N} \gamma^{\mu} e) (\bar{q} u) D_{\mu} H $ & 
        $\opQ{HlNu^2D}$ & 
        $ (\bar{u} \gamma^{\mu} u) (\bar{l} N) D_{\mu} \tilde{H} $ \\

        $\opQ{HelN^2D}$ &
        $ (\bar{N} \gamma^{\mu} N) (\bar{l} e) D_{\mu} H $ &
        $\opQ{HdN^2qD}$ &
        $ (\bar{N} \gamma^{\mu} N) (\bar{q} d) D_{\mu} H $ \\
        
        $\opQ{HN^2quD}$ &
        $ (\bar{N} \gamma^{\mu} N) (\bar{q} u) D_{\mu} \tilde{H} $ &
        $\opQ{HlN^3D}$& 
        $ (\bar{N} \gamma^{\mu} N) (\bar{l} D_{\mu} N)\tilde{H}  $ \\
        
        \hline
        \hline
        $\opQ{Hd^2NQD}$ & 
        \blvmc{$ \epsilon_{\alpha \beta \gamma} \epsilon_{jk} (Q_{L}^{j \alpha} C \gamma^{\mu} d_{R}^{\beta})(d_{R}^{\gamma} C N_{R}) D_{\mu} H^{k} $  }\\
        
        $\opQ{HNq^3D}$& 
        \blvmc{$ \epsilon_{\alpha \beta \gamma} \epsilon_{mn} \epsilon_{jk} (q_{L}^{j \alpha} C \gamma^{\mu} N_{R}) (D_{\mu} q_{L}^{k \beta} C q_{L}^{m \gamma}) \tilde{H^{n}} $ } \\
        
        $\opQ{HdNQD}$ & 
        \blvmc{$\epsilon_{\alpha \beta \gamma} \epsilon_{jk} (Q_{L}^{j \alpha} C \gamma^{\mu} u_{R}^{\beta}) (d_{R}^{\gamma} C N_{R}) D_{\mu} \tilde{H^{k}} $ } \\
        
        $\opQ{HlN^3D}$ &         
        \blvmc{$ \epsilon_{jk} (N_{R} C \gamma^{\mu} l_{L}^{j}) (N_{R} C N_{R}) D_{\mu} H^{k} $} \\
        \hline
        \end{tabular}}
	\caption{Operators in $\mathbf{\Psi^4 H D }$ class. $(+\hc)$ denotes the hermitian conjugate of that operator. The operators listed in the lower section of the table violate Baryon and/or Lepton numbers. }
	\label{tab:dim8-ops-4}
\end{table}

In Table~\ref{tab:counting} we present the class-wise counting of operators in $\nsmeft$ at D8. To maintain generality, we provide the counting for an arbitrary number of fermion generations, $n_i$, where $i \in \{q, l, u, d, e, N\}$ denotes the fermion species. We further specify the ``type" for each class, defined as the specific combination of fields and covariant derivatives that constitute a gauge-invariant term. Finally, by setting the number of generations to $n_g = 1$ and $n_g = 3$, \footnote{Setting $n_g =1,3$ means setting all $n_i$s to 1,3.} we derive the total number of independent operators. These results have been cross-checked and are in full agreement with the literature \cite{Li:2021tsq}.

\begin{table}[h]
    \centering
    \renewcommand{\arraystretch}{1.4}
    {\scriptsize
    \begin{tabular}{c|c|c|c|c|c}
    Class & $(\Delta B, \Delta L)$ & no. of operators & $\mathcal{N}_\text{type}$ & $n_g=1$ & $n_g=3$ \\
    \hline
    \hline
    $\mathbf{\Psi^2 H^5}$ &$(0,0)$& $2\times(n_l n_N ) $ & $2\times1=2$  & 2 & 18 \\
    \hline
    $\mathbf{\Psi^2H^4 D} $ &$(0,0)$& $ 2\times (n_{e} n_{N}) + n_N^2 $ &$2\times1+1=3$& 3 & 27 \\
    \hline
    $\mathbf{\Psi^2 H^3 X}$ && $2\times(3\;n_l n_N ) $ & $2\times2 = 4$& 6 & 54  \\
    \hline
    $\mathbf{\Psi^2 H^3 D^2}$ &$(0,0)$& $2\times(6\;n_l n_N ) $& $2\times1=2$& 12 & 108 \\
    \hline
    $\mathbf{\Psi^2 H^2 X D}$ &$(0,0)$& $2\times((4\;n_l n_N ) + 4\; n_N^2) $& $2\times6=12$& 16 & 144\\
    \hline
    $\mathbf{\Psi^2 H^2 D^3}$ &$(0,0)$& $2\times(n_e n_N ) + 2\;n_N^2 $ &$2\times1+1=3$& 4 & 36 \\
    \hline
    $\mathbf{\Psi^2 H X^2}$ &$(0,0)$& $2\times(10\;n_l n_N )$ &$2\times8=16$& 20 & 180 \\
    \hline
    $\mathbf{\Psi^2 X^2 D}$ &$(0,0)$& $ 3\;n_N^2 $ & $3$ & 3  & 27 \\
    \hline
    $\mathbf{\Psi^2 H X D^2}$ &$(0,0)$& $2\times(6\;n_l n_N )$& $2\times4=8$ & 12 & 108  \\
    \hline
    
    $\mathbf{\Psi^4 H^2}$ &$(0,0)$& $ \frac{1}{4} n_N \Big(n_N \left(4 n_d^2+4 n_e^2+8 n_q^2+4 n_u^2+1\right)+$ & $2\times14+6=34$&  44 &  3423\\
    && $ 8 n_e \left(n_d n_u+n_q^2\right)+4 n_l (10 n_d n_q+8 n_q n_u+1)+  $ &  & & \\
    && $ 12 n_l^2 (2 n_e+n_N)+n_N^3+2 n_N^2\Big) $&& &  \\
    \cline{2-3}
    &$(\pm1,\mp1)$ & $2\times \Big( n_d^2 n_N n_u + n_d n_N n_q^2+\frac{1}{2} n_N (n_q-1) n_q n_u \Big) $ & && \\
    \cline{2-3}
    &$(0,\pm4)$&
    $ 2\times\frac{1}{12} n_N (n_N+1) \Big(3 n_l^2+3 n_l+(n_N-1) n_N \Big)$& & & \\
    \hline
    
    $\mathbf{\Psi^4 D^2}$ & $(0,0)$ & 
    $\frac{1}{2} n_N \Big(n_N \left(4 n_d^2+4 n_e^2+4 n_l^2+4 n_q^2+4 n_u^2+1\right)+$& $2\times7+6=20$& 33 & 2625  \\
    &&$ 8 n_l n_q n_u+ n_N^3\Big) $ && & \\
    \cline{2-3}
    & $(\pm1,\mp1)$ & $2\times\frac{1}{2} n_N \Big(n_e \left(4 n_d n_u+3 n_l^2-n_l\right)+$&&  & \\
    && $ n_d \left(3 n_d n_u+6 n_l n_q+2 n_q^2-n_u\right)\Big) $ &&& \\
    \cline{2-3}
    &$(0,\pm4)$& $2\times\frac{1}{8} n_N( n_N+1) \left(n_N^2+n_N+2\right)$& & & \\
    \hline
    
    $\mathbf{\Psi^4 X}$ & $(0,0)$ &
    $2\times\frac{1}{4} n_N \Big(n_N \left(8 n_d^2+8 n_l^2+12 n_q^2+8 n_u^2-1\right)+ $&$2\times35=70$&  82  &  6630\\
    && $ 4 n_e \left(4 n_d n_u+3 n_l^2\right) + $ & && \\
    && $ 12 n_l n_q (3 n_d+2 n_u)+4 n_e^2 n_N+n_N^3\Big) $&& & \\
    \cline{2-3}
    &$(\pm1,\mp1)$&
    $2\times (\frac{9}{2} n_d^2 n_N n_u+ 4 n_d n_N n_q^2+\frac12n_d n_N n_u)$ && &  \\
    \cline{2-3}
    &$(0,\pm4)$&
    $2\times\frac{n_N}{8} (n_N-2) (n_N-1)  (n_N+1)$&& & \\
    \hline

    $\mathbf{\Psi^4 HD}$ & $(0,0)$ &
    $ n_N \Big(6 n_d^2 n_l+6 n_d (n_e n_q+n_l n_u+n_N n_q)+6 n_e^2 n_l+ $ & $2\times16=32$&  86 &  7020\\
    && $6 n_e (n_l n_N+n_q n_u)+6 n_l^3+3 n_l n_N^2-n_l n_N+$ &&& \\
    && $ 12 n_l n_q^2+6 n_l n_u^2+6 n_N n_q n_u\Big) $ &&& \\
    \cline{2-3}
    & $(\pm1,\mp1)$ & $2\times\frac{1}{2} n_N n_q \left(3 n_d^2+6 n_d n_u+n_d+2 n_q^2\right)$ & && \\
    \cline{2-3}
    & $(0,\pm4)$ & $2\times \frac{1}{2} n_l (n_N-1) n_N^2 $ & && \\
    \hline
    \hline
    \textbf{Total} & & & 209 &323  &20400 \\
    \hline
    \end{tabular}}
	\caption{Summary of all the operator classes presented above. The last three columns indicate the number of operators in each class.}
	\label{tab:counting}
\end{table}
\clearpage
\newpage
\subsection{Possible Future Directions}

Upon construction of the complete operator set, several other potential phenomenological features become apparent. The corresponding operators and signatures are discussed below.

\begin{itemize}

\item \textbf{\(\psi^4 H^2\):}  
Decay channels such as \(H \to \mu^+ \mu^- N N\), \(H \to N N N N\), 
\(H \to \ell \ell \ell N\), \(H \to \text{MET} + \ell \ell N\), and 
\(H \to j j \ell N\) provide distinctive experimental signatures at the 
HL-LHC and at future muon colliders. In lepton-number--violating 
scenarios, additional modes like \(H \to j j j N\) can also arise. 
For high-energy colliders, we consider a benchmark cutoff scale of 
\(\Lambda = 1\)~TeV, which implies significant suppression of these rates. 
At the FCC-hh, complementary channels, such as 
\(pp \to N + \gamma + \ell\), become accessible and offer further 
discovery potential.

\item \textbf{\(\psi^4 H D\):}  
This class of operators can give rise to a variety of distinctive final states at both hadron and lepton colliders. For instance, at the HL-LHC one may encounter processes such as \(pp \to h \to \text{MET} + N N N + Z\) or \(pp \to Z/\gamma \to \text{MET} + N N N + h\). Similarly, at future muon colliders, analogous signatures can arise through \(\mu^+ \mu^- \to h \to \text{MET} + N N N + Z\) and \(\mu^+ \mu^- \to Z/\gamma \to \text{MET} + N N N + h\). These channels highlight the complementarity between hadronic and leptonic machines in probing such operators.  

\end{itemize}
These potential phenomenological studies are outlined here for completeness but are postponed for dedicated future study.

\bibliographystyle{JHEP}
\bibliography{ref.bib}

\end{document}